\documentclass[a4paper,twocolumn,10pt,accepted=2026-02-23]{quantumarticle}

\pdfoutput=1
\usepackage[utf8]{inputenc}
\usepackage[english]{babel}

\usepackage{graphicx}
\usepackage{epstopdf}
\usepackage{amsmath}
\usepackage{amssymb}
\usepackage{mathrsfs}
\usepackage{amsthm}

\usepackage{mathtools,mleftright}
\mleftright
\usepackage{bm}
\usepackage{bbm}
\usepackage{url}
\usepackage[T1]{fontenc}
\usepackage{csquotes}
\MakeOuterQuote{"}
\usepackage{multirow}
\usepackage{makecell}
\usepackage[square,numbers,sort&compress]{natbib}
\usepackage{hyperref}

\newtheoremstyle{note}
  {\topsep/2}               
  {\topsep/2}               
  {}                      
  {\parindent}            
  {\itshape}              
  {.}                     
  {5pt plus 1pt minus 1pt}
  {}

\theoremstyle{note}
\newtheorem{theorem}{Theorem}

\theoremstyle{definition}

\theoremstyle{remark}

\newcommand{\equad}{\,\hphantom{=}\,}
\newcommand{\lsp}{\hspace{0.1em}}

\newcommand{\tr}{\operatorname{tr}}

\newcommand{\bP}{\bar{P}}

 \newcommand{\rme}{\mathrm{e}}
 
 \newcommand{\rmi}{\mathrm{i}}

 \newcommand{\rmD}{\mathrm{D}}

 \newcommand{\rmG}{\mathrm{G}}

 \newcommand{\rmW}{\mathrm{W}}

 \newcommand{\SD}{\mathrm{SD}}
 \newcommand{\CSD}{\mathrm{CSD}}
 \newcommand{\MUB}{\mathrm{MUB}}
 \newcommand{\SMUB}{\mathrm{SMUB}}

 \newcommand{\caH}{\mathcal{H}}

 \newcommand{\caO}{\mathcal{O}}
 \newcommand{\caP}{\mathcal{P}}

 \newcommand{\caS}{\mathcal{S}}
 
 \newcommand{\caZ}{\mathcal{Z}}

 \newcommand{\bbZ}{\mathbb{Z}}

 \newcommand{\id}{1}

 \newcommand{\bfj}{\mathbf{j}}

 \newcommand{\bfm}{\mathbf{m}}
 \newcommand{\bfn}{\mathbf{n}}
 \newcommand{\bfo}{\mathbf{o}}

 \newcommand{\scrB}{\mathscr{B}}

\newcommand{\ta}{\tilde{a}}
\newcommand{\tn}{\tilde{n}}
\newcommand{\tB}{\tilde{B}}
\newcommand{\tP}{\tilde{P}}
\newcommand{\tOmega}{\tilde{\Omega}}

 \newcommand{\bmu}{\bm{u}}
 \newcommand{\bmv}{\bm{v}}

\newcommand{\be}{\begin{equation}}
\newcommand{\ee}{\end{equation}}
\newcommand{\ba}{\begin{align}}
\newcommand{\ea}{\end{align}}

\def\<{\langle}  
\def\>{\rangle}  

\def\eqref#1{\textup{(\ref{#1})}}  
\newcommand{\eref}[1]{Eq.~\textup{(\ref{#1})}}

\newcommand{\fref}[1]{Fig.~\ref{#1}}
\newcommand{\Fref}[1]{Figure~\ref{#1}}
\newcommand{\fsref}[1]{Figs.~\ref{#1}}

\newcommand{\tref}[1]{Table~\ref{#1}}
\newcommand{\Tref}[1]{Table~\ref{#1}}
\newcommand{\tsref}[1]{Tables~\ref{#1}}

\newcommand{\sref}[1]{Sec.~\ref{#1}}

\newcommand{\thref}[1]{Theorem~\ref{#1}}
\newcommand{\Thref}[1]{Theorem~\ref{#1}}

\newcommand{\cref}[1]{Conjecture~\ref{#1}}
\newcommand{\Cref}[1]{Conjecture~\ref{#1}}

\newcommand{\aref}[1]{Appendix~\ref{#1}}
\newcommand{\asref}[1]{Appendices~\ref{#1}}

\newcommand{\rcite}[1]{Ref.~\cite{#1}}
\newcommand{\rscite}[1]{Refs.~\cite{#1}}

\begin{document}

\title[Article Title]{Universal and Efficient Quantum State Verification via Schmidt Decomposition and Mutually Unbiased Bases}

\author{Yunting Li}
\affiliation{State Key Laboratory of Surface Physics, Department of Physics, and Center for Field Theory and Particle Physics, Fudan University, Shanghai 200433, China}
\affiliation{Institute for Nanoelectronic Devices and Quantum Computing, Fudan University, Shanghai 200433, China}
\affiliation{Shanghai Research Center for Quantum Sciences, Shanghai 201315, China}

\author{Huangjun Zhu}
\email{zhuhuangjun@fudan.edu.cn}
\orcid{0000-0001-7257-0764}
\affiliation{State Key Laboratory of Surface Physics, Department of Physics, and Center for Field Theory and Particle Physics, Fudan University, Shanghai 200433, China}
\affiliation{Institute for Nanoelectronic Devices and Quantum Computing, Fudan University, Shanghai 200433, China}
\affiliation{Shanghai Research Center for Quantum Sciences, Shanghai 201315, China}

\begin{abstract}
Efficient verification of multipartite quantum states is crucial to many  applications in quantum information processing. By virtue of Schmidt decomposition and mutually unbiased bases, here we propose a universal protocol to verify arbitrary multipartite pure quantum states using adaptive local projective measurements. Moreover, we establish a universal upper bound on the sample complexity that is independent of the local dimensions. Numerical calculations further indicate that Haar-random pure states can be verified with a constant sample cost, irrespective of the qudit number and local dimensions, even in the adversarial scenario in which the source cannot be trusted. As alternatives, we provide several simpler variants that can achieve similar high efficiencies without using  Schmidt decomposition. The simplest variant consists of only two distinct tests.
\end{abstract}

\maketitle
\tableofcontents

\section{Introduction}

Multipartite entangled states play central roles in quantum information processing \cite{Horodecki2009,GUHNE2009,Walter2016}. For many practical applications, it is crucial to verify whether the actual state prepared is sufficiently close to the ideal target state with limited resources. However, traditional methods, such as quantum state tomography, are too resource-intensive.
Recently, a powerful approach known as \emph{quantum state verification} (QSV) has attracted intensive attention \cite{CarrEKK21,MorrSGD22,YuSG22,HayaMT06,AoliGKE15,HangKSE17,TakeM18,PallLM18,Zhu2019Efficient,Zhu2019General,WuBGL21,LiuSHZ21,GocaSD22}. So far, efficient  protocols based on local projective measurements have been constructed for bipartite pure states \cite{HayaMT06,Haya09G,PallLM18,ZhuH19O,WangH19,LiHZ19,YuSG19}, stabilizer states [including graph states and Greenberger-Horne-Zeilinger (GHZ) states in particular] \cite{HayaM15,FujiH17,HayaH18, PallLM18,Markham2018ASP,Zhu2019Efficient,Zhu2019General,Li2020GHZ}, hypergraph states \cite{ZhuH19E}, Dicke states \cite{LiuYSZ19,Li2021Dicke}, and ground states of frustration-free Hamiltonians (including Affleck-Kennedy-Lieb-Tasaki states) \cite{HangKSE17,LiYT2023AKLT,Zhu2024FFH}. In addition, several verification protocols have been successfully demonstrated in experiments \cite{ZhanZCP20,ZhanLYP20,LuXCC20,JianWQC20}.

Unfortunately, most verification protocols known so far are tailored to quantum states with special structures and thus have  limited scopes of applications. As a major breakthrough, recently, Huang, Preskill, and Soleimanifar (HPS) proposed a protocol that can certify almost all $n$-qubit pure states using single-qubit measurements~\cite{huang2025certifying}. In addition, \rcite{Liu2023Homo} proposed a general method for constructing homogeneous verification strategies. Nevertheless, as far as we know, no explicit protocol based on local measurements is known in the literature that can verify all $n$-qudit pure states directly.
The efficiency limit of local  measurements in QSV is still quite elusive.

In this work, we propose a universal protocol to verify arbitrary multipartite pure quantum states using adaptive local projective measurements. Our protocol is based on two elementary concepts in quantum information theory, namely, \emph{Schmidt decomposition} (SD) \cite{NielC10book} and \emph{mutually unbiased bases} (MUB) \cite{Schwinger1960,Ivonovic1981,WOOTTERS1989363,DURT2010}, and is  referred to as the SD protocol henceforth. While the importance of SD and MUB in QSV was recognized  before \cite{ZhuH19O,LiHZ19,YuSG19,Li2020GHZ},  little is known about their potential in verifying generic multipartite states before this work.
We prove a universal upper bound on the sample complexity of the SD protocol, which increases exponentially with the qudit number $n$, but is independent of the local dimensions. Moreover, numerical calculations indicate that Haar-random pure states can be verified with a constant sample cost, irrespective of the qudit number and local dimensions. This result holds even if the source cannot be trusted, which is typical in the adversarial scenario, such as in blind measurement-based quantum computation \cite{RausB01,BroaFK09,HayaM15,Fitzsimons2017,LiZH23}.

To reduce the computational complexity, we also introduce a simpler verification protocol based on MUB, henceforth referred to as the MUB protocol, which avoids SD and is more amenable to experimental realization. 
Surprisingly, the MUB protocol can also be applied to arbitrary multipartite quantum systems, and many variants  can achieve comparable performance to the SD protocol even if only a constant number of distinct tests are accessible. Notably, in the simplest variant, only two bases are required for each party except for the last party (a complete set of MUB is not necessary), and only two distinct tests in total are required. This intriguing phenomenon may reflect certain hidden characteristics of generic multipartite pure states yet to be unraveled.  Our work indicates that all multipartite pure states can be verified efficiently using local projective measurements, and most pure states can be verified using a constant sample cost and a constant number of distinct tests.

\section{Quantum state verification}

Consider a quantum device that is supposed to produce the target state $|\Psi \rangle$ in the Hilbert space $\caH$. In practice, it actually produces $N$ states $\rho_1,\rho_2,\dots,\rho_N$ in $N$ runs. To verify whether these states are sufficiently close to the target state on average, we can perform a random test from a given set of accessible tests. Each test is tied  to a two-outcome measurement  $\{E_m, \id-E_m\}$, where the \emph{test operator} $E_m$ means passing the test and the number 1 also denotes the identity operator. To guarantee that the target state $|\Psi\>$ can always pass each test, the operator $E_m$ should satisfy the condition $ E_m | \Psi \rangle =|\Psi \rangle$.

Suppose the test  $E_m$ is performed with probability $p_m$; then the performance of the verification strategy is determined by the \emph{verification operator}  $\Omega = \sum_m p_m E_m$. If $\rho$ is a quantum state on $\caH$ whose fidelity with the target state is at most $1-\varepsilon$, that is, $\langle \Psi |\rho |\Psi \rangle \leq 1-\varepsilon$; then the maximum probability that $\rho$ can pass each test on average reads \cite{PallLM18,Zhu2019Efficient,Zhu2019General}:
\begin{equation}
\max_{\langle \Psi |\rho |\Psi \rangle \leq 1-\varepsilon} \tr(\Omega \rho)=1-[1-\beta(\Omega)]\varepsilon 
= 1-\nu(\Omega)\varepsilon.
\end{equation}
Here $\beta(\Omega)$ denotes the second largest eigenvalue of $\Omega$, and $\nu(\Omega)=1-\beta(\Omega)$ denotes the \emph{spectral gap} from the largest eigenvalue.
Let $\bar{E}_m = E_m - |\Psi\>\<\Psi|$ and $\bar{\Omega} = \Omega - |\Psi\>\<\Psi|= \sum_m p_m \bar{E}_m$; then $\beta(\Omega) = \|\bar{\Omega} \|$.
Let $\varepsilon_r = 1-\< \Psi | \rho_r |\Psi \>$ be the infidelity of the state $\rho_r$ prepared  in the $r$-th run and $\bar{\varepsilon} = \sum_{r=1}^N \varepsilon_r /N$ the average infidelity. 
Then the probability that the states  $\rho_1,\rho_2,\dots,\rho_N$  can pass all $N$ tests satisfies
$\prod_{r=1}^N [1-\nu(\Omega)\varepsilon_r] \le [1-\nu(\Omega)\bar{\varepsilon}]^N$.
To verify the target state within infidelity $\varepsilon$ and signiﬁcance level $\delta$, which means $[1-\nu(\Omega)\varepsilon]^N\leq \delta$, 
it suffices to take
\cite{PallLM18,Zhu2019Efficient,Zhu2019General}
\begin{equation}
N=\left \lceil \frac{\ln \delta}{\ln [1-\nu(\Omega)\varepsilon]} \right \rceil 
\le \left \lceil \frac{\ln (\delta^{-1})}{\nu(\Omega)\varepsilon} \right \rceil ,
\end{equation}
which is inversely proportional to $\nu(\Omega)$.

To achieve high efficiency, we need to construct a verification operator $\Omega$ with a large spectral gap, which means the second largest eigenvalue $\beta(\Omega)$ is small. If there is no restriction on the measurements, then $\nu(\Omega)$ can attain the maximum value 1 when $\Omega=|\Psi\>\<\Psi|$. However, it is in general extremely difficult to realize this strategy, which relies on an entangling measurement.  
It is also  highly nontrivial to construct an efficient verification strategy  using local projective measurements if the target state $|\Psi\>$ lacks a simple structure, which is the main motivation behind the current work.

If  the set of test operators $\{E_m\}_m$ is fixed, then the minimum of $\beta(\Omega)$ and the optimal  probabilities $p_m$ can be determined via semidefinite programming (SDP) \cite{vandenberghe1996semidefinite}:
\begin{equation}
\begin{aligned}
\!&\mathrm{minimize}  &&f,  \\
\!&\mathrm{subject\: to}  &&\sum_m p_m \bar{E}_m \le f, \, p_m \ge 0, \, \sum_m p_m =1. 
\end{aligned} \label{eq:sdp}
\end{equation}

\section{Schmidt-decomposition protocol}

\subsection{Protocol and performance} \label{sec:SDpp}

\begin{figure}[tbp]
\centering
\includegraphics[width=0.45\textwidth]{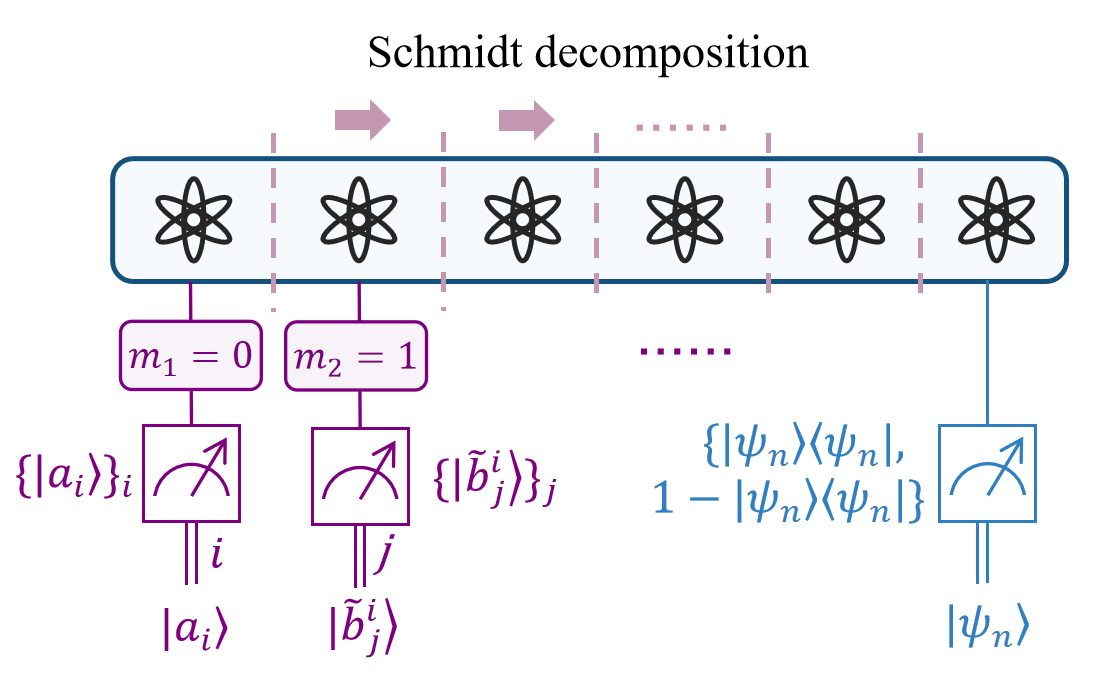}
\vspace{-1ex}
\caption{Schematic of the SD protocol for verifying an $n$-qudit state $|\Psi\>$. The first party performs a projective measurement in either the Schmidt basis $(m_1=0)$ or a MUB $(m_1=1)$ with respect to the Schmidt basis. Then party $k$ for each $k=2,3,\ldots,n-1$ performs a projective measurement in succession in either the Schmidt basis $(m_k=0)$ or a MUB $(m_k=1)$ associated with the conditional reduced state of $|\Psi\>$ for parties $k,k+1,\ldots, n$ depending on the previous measurement choices and outcomes. Finally,  party $n$ performs the projective measurement $\{|\psi_n\>\<\psi_n|,1-|\psi_n\>\<\psi_n|\}$, where  
$|\psi_n\>$ is  the conditional reduced state of $|\Psi\>$ for party $n$, and the first outcome means passing the test. }
\label{fig:SchmidtFramework}
\end{figure}

Here we propose a universal verification protocol based on SD and MUB, which requires only adaptive local projective measurements, as illustrated in \fref{fig:SchmidtFramework}. For simplicity, we  focus on an $n$-qudit system with  Hilbert space $\caH=\caH_d^{\otimes n}$ of dimension $d^n$, but generalization to systems of different local dimensions is immediate. The basis states in any basis on $\caH_d$ can be labeled by the ring $\bbZ_d$  of integers modulo $d$. Recall that two (orthonormal) bases $\{|\phi_i\>\}_{i \in \bbZ_d}$ and $\{|\varphi_j\>\}_{j \in \bbZ_d}$ on $\caH_d$ are mutually unbiased if $|\<\phi_i|\varphi_j\>|=1/\sqrt{d}$ for all $i, j\in \bbZ_d$ \cite{Schwinger1960,Ivonovic1981,WOOTTERS1989363,DURT2010}. MUB have numerous applications, including quantum state estimation \cite{Ivonovic1981,WOOTTERS1989363,Roy2007,Adamson2010,Zhu2014,yan2024experimental},  quantum verification \cite{
ZhuH19O,LiHZ19,YuSG19,Li2020GHZ}, and entanglement detection \cite{TothG05,Bavaresco2018,Bae2022}.

Let $|\Psi\> = \sum_{i\in \bbZ_d} s_i |a_i\> |B_i\>$ be a  SD of the target state $|\Psi\>$ with respect to  the first party and the remaining $n-1$ parties,
where $s_i$ are Schmidt coefficients, $\{|a_i\>\}_{i\in \bbZ_d}$ forms an orthonormal basis in $\caH_d$, and $\{|B_i\>\}_{i\in \bbZ_d}$ consists of orthonormal states in $\caH_d^{\otimes (n-1)}$. 
The first party can perform a projective measurement in either the Schmidt basis $\{|a_i\>\}_{i \in \bbZ_d}$ or a MUB with respect to the Schmidt basis, say,  the Fourier basis $\{|\tilde{a}_i\>\}_{i\in \bbZ_d}$ with $|\tilde{a}_i\> = \sum_{j\in \bbZ_d} \omega^{ij} |a_j\> / \sqrt{d}$ and $\omega = \rme^{2\pi \rmi /d}$. Given the outcome $i$, the normalized reduced state of the remaining parties is either $|B_i\>$ or $|\tilde{B}_i\> = \sum_{j\in \bbZ_d} \omega^{-ij} s_j |B_j\>$ depending on the measurement basis of the first party.
The protocol then recursively applies this procedure to the conditional reduced state of the remaining parties: each party (except for the last party) performs a projective measurement in either the  Schmidt basis or a  MUB; then the last party performs the projective measurement $\{|\psi_n\>\<\psi_n|,1-|\psi_n\>\<\psi_n|\}$, where  
$|\psi_n\>$ is  the conditional reduced state of $|\Psi\>$ (depending on the measurement choices and outcomes of the previous parties). The test is passed if the last party obtains the first outcome.

The above verification protocol consists of $2^{n-1}$ tests, corresponding to the $2^{n-1}$ basis choices of the first $n-1$ parties. Each test (projector) can be labeled by a string  $\bfm=(m_1, m_2, \dots, m_{n-1})$ in $\bbZ_2^{n-1}$, where $m_k=0$ denotes the Schmidt basis and $m_k=1$ denotes the MUB.
When $n=2$, so that $|\Psi\>$ is a bipartite pure state,   the two test projectors read
\begin{equation}
\begin{aligned}
    P_{0} &= \sum_{i\in \bbZ_d\mid s_i>0} |a_i\> \<a_i| \otimes |B_i\>\<B_i|, \\
    P_{1} &= \sum_{i\in \bbZ_d} |\tilde{a}_i\> \<\tilde{a}_i| \otimes |\tilde{B}_i\>\<\tilde{B}_i|,
\end{aligned}  \label{eq:P0P1}
\end{equation}
which reproduce  the results in \rcite{LiHZ19}. When $n=3$, explicit expressions for the four test projectors can be found in \sref{sec:ThreeQuditTests}. If the test $P_\bfm$ is performed with probability $p_\bfm$, then the verification operator reads $\Omega_{\SD} = \sum_{\bfm\in \bbZ_2^{n-1}} p_\bfm P_\bfm$. The following theorem establishes a universal lower bound (upper bound) for the spectral gap (sample complexity) of the SD protocol based on uniform probability, that is, $p_\bfm=2^{1-n}$ for  $\bfm\in \bbZ_2^{n-1}$; see \sref{sec:thmQSVSDgapProof} for a proof. 

\begin{theorem}\label{thm:QSVSDgap}
	Suppose $|\Psi\>\in \caH$ is an $n$-qudit pure state with $n\geq 2$, and $\Omega_\SD$ is the verification operator of $|\Psi\>$ tied to the SD protocol with uniform probability. Then $\nu(\Omega_\SD)\geq 2^{1-n}$ and the number of tests required to verify $|\Psi\>$ within infidelity $\varepsilon$ and significance level $\delta$ satisfies
\begin{equation}
	N\le \left \lceil 2^{n-1}\varepsilon^{-1}\ln \left(\delta^{-1}\right)
\right\rceil.  \label{eq:QSVSDNUB}
\end{equation}
\end{theorem}

\Thref{thm:QSVSDgap} shows that the SD protocol is a universal verification protocol that applies to arbitrary multipartite pure states, which is the first verification protocol with this capability as far as we know. Although the lower bound for $\nu(\Omega_\SD)$ decreases exponentially with $n$, for most pure states, the spectral gaps  $\nu(\Omega_\SD)$ are larger than the universal constant $1/5$ (independent of $d$ and $n$) according to extensive numerical simulations. We still do not have a simple explanation for this universal constant. Nevertheless, this result shows that  Haar-random pure states can be verified much more efficiently than the worst-case guarantee in \thref{thm:QSVSDgap}. We hope that our observation can stimulate further progress on this topic.

\begin{figure}[tb]
	\centering
	\includegraphics[width=0.45\textwidth]{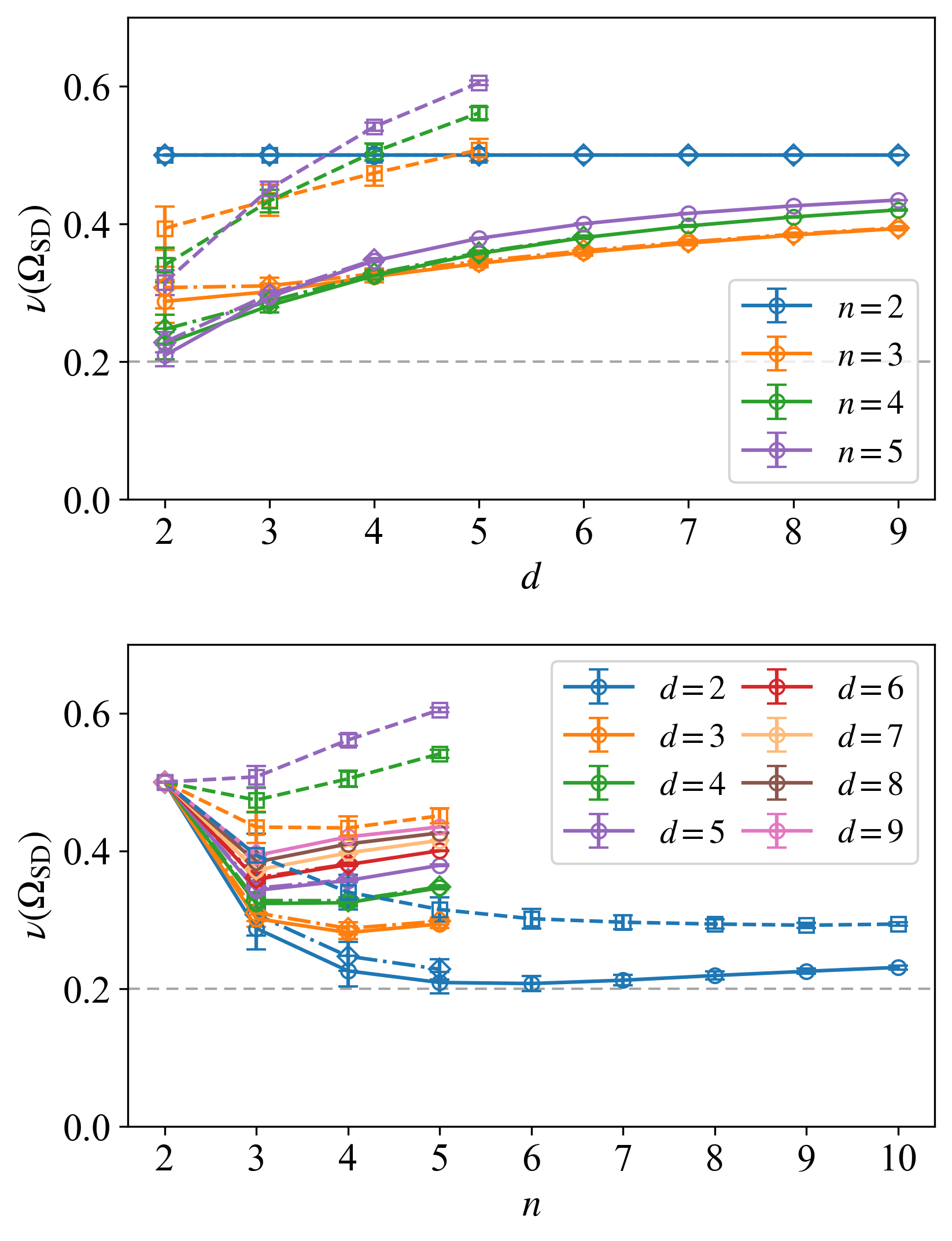}
	\caption{Average spectral gaps achieved by the SD and CSD protocols for $n$-qudit Haar-random pure states (see \tref{tab:SampleNumHaar} in \aref{app:SampleNumHaar}  for the numbers of states sampled). Solid lines with circles (dashed lines with squares) represent the SD (CSD) protocol under the uniform probability distribution, while
    dash-dot lines with diamonds represent the SD protocol under the optimized probability distribution determined by SDP. The gray dashed lines correspond to the spectral gap of $1/5$.}
	\label{fig:HaarSD}
\end{figure}

\Fref{fig:HaarSD} shows the average spectral gap for  $n$-qudit Haar-random pure states as a function of $d$ and $n$. Surprisingly,  the average spectral gap increases monotonically with the local dimension $d$, so it is easier to verify quantum states with higher local dimensions. For a given $d$, the spectral gap first decreases but eventually increases with $n$; the turning point $n^*$ for $n$ is usually smaller than~5, except when $d=2$. In addition, the spectral gap  can usually be increased by a few percent if we optimize test probabilities via SDP in \eref{eq:sdp}. Results on typical multipartite states, including GHZ states and Dicke states (see \fref{fig:DickeGapn}),  are shown in
\aref{app:SDprotocol}. Note that probability optimization can significantly increase the spectral gaps for  GHZ states, although its efficacy for generic states is limited.

\subsection{Test projectors featured in the SD protocol for a three-qudit pure state} \label{sec:ThreeQuditTests}

For concreteness, here we provide explicit expressions for the four test projectors in the case $n=3$. Suppose $|\Psi\>\in \caH=\caH_d^{\otimes 3}$ is a three-qudit pure state to be verified, where $\caH_d$ is a $d$-dimensional Hilbert space. Let $|\Psi\> = \sum_{i\in \bbZ_d} s_i |a_i\> |B_i\>$ be the SD of $|\Psi\>$ between the first and  remaining parties, where
$\{|a_i\>\}_{i\in \bbZ_d}$ forms an orthonormal basis on  $\caH_d$, and $\{|B_i\>\}_{i\in \bbZ_d}$ consists of $d$ orthonormal states in $\caH_d^{\otimes 2}$. Note that our main results are independent of the specific SD chosen. Let 
\begin{equation}
    |B_i\> = \sum_{j\in \bbZ_d} t_j^i |b_j^i\> |c_j^i\> 
\end{equation}
be the SD of $|B_i\>$, 
where $t_j^i$ are the Schmidt coefficients satisfying $\sum_{j\in \bbZ_d} \left(t_j^i\right)^2 = 1$, while $\{|b_j^i\>\}_{j \in \bbZ_d}$ and $\{|c_j^i\>\}_{j \in \bbZ_d}$ are the Schmidt bases for the second and third parties, respectively. Then the target state can be expressed as
\begin{equation}
    |\Psi\> =\sum_{i\in \bbZ_d} s_i |a_i\> |B_i\> 
	=\sum_{i\in \bbZ_d}\sum_{j\in \bbZ_d} s_i t_j^i |a_i\>  |b_j^i\> |c_j^i\>.
\end{equation}

Suppose the first party performs the projective measurement in the Schmidt basis $\{|a_i\>\}_{i \in \bbZ_d}$ and obtains outcome $i$. Then the reduced state of the remaining parties is $|B_i\>$. Now, if the second party performs the projective measurement in the Schmidt basis $\{|b_j^i\>\}_{j \in \bbZ_d}$ and obtains outcome $j$, then the reduced state of the third party is $|c_j^i\>$. Finally, the third party can perform the binary measurement $\{|c_j^i\>\<c_j^i|, 1-|c_j^i\>\<c_j^i|\}$ to single out the first outcome. Equivalently, the third party can perform the projective measurement in the Schmidt basis $\{|c_j^i\>\}_{j \in \bbZ_d}$. The resulting test projector, labeled by the string $\bfm=(0,0)\in \bbZ_2^2$, reads
\begin{equation}
    P_{(0,0)} = \sum_{i\mid  s_i>0} \sum_{j\mid  t_j^i>0} |a_i\> \<a_i| \otimes |b_j^i\> \<b_j^i| \otimes |c_j^i\> \<c_j^i|.  \label{eq:P00}
\end{equation}
Alternatively,  the second party can perform the projective measurement in a basis that is mutually unbiased with $\{|b_j^i\>\}_{j \in \bbZ_d}$. For concreteness, here we shall focus on the Fourier basis with respect to the Schmidt basis:
\begin{equation}
    |\tilde{b}_j^i\> = \sum_{l\in \bbZ_d} \frac{1}{\sqrt{d}} \omega^{jl} |b_l^i\>,
\end{equation}
where $\omega=\rme^{2\pi\rmi/d}$, 
but other MUB are equally good. If the second party obtains outcome $j$, then the reduced state of the third party reads
\begin{equation}
    |\tilde{c}_j^i\> = \sum_{l\in \bbZ_d} t_l^i \omega^{-jl} |c_l^i\>,
\end{equation}
which can be singled out by a suitable binary projective measurement. The resulting test projector, labeled by the string $\bfm=(0,1)$,  reads
\begin{equation}
    P_{(0,1)} = \sum_{i\in \bbZ_d \mid  s_i>0} \sum_{j\in \bbZ_d} |a_i\> \<a_i| \otimes |\tilde{b}_j^i\> \<\tilde{b}_j^i| \otimes |\tilde{c}_j^i\> \<\tilde{c}_j^i|. \label{eq:P01}
\end{equation}

To introduce the remaining two test projectors, let $\{|\tilde{a}_i\>\}_{i\in \bbZ_d}$ with $|\tilde{a}_i\> = \sum_{j\in \bbZ_d} \omega^{ij} |a_j\> / \sqrt{d}$ be the Fourier basis  with respect to the Schmidt basis $\{|a_i\>\}_{i \in \bbZ_d}$. Let 
\begin{equation}
		|\tilde{B}_i\> = \sum_{j\in \bbZ_d} \omega^{-ij} s_j |B_j\> 
		= \sum_{j\in \bbZ_d} \omega^{-ij} s_j \sum_{l\in \bbZ_d} t_l^j |b_l^j\> |c_l^j\>
\end{equation}
for $i\in \bbZ_d$ and construct SD as follows:
\begin{equation}
    |\tilde{B}_i\> = \sum_{j\in \bbZ_d} u_j^i |\bar{b}_j^i\> |\bar{c}_j^i\>,
\end{equation}
where  $u_j^i$ are the Schmidt coefficients of $|\tilde{B}_i\>$, while  $\{|\bar{b}_j^i\>\}_{j \in \bbZ_d}$ and $\{|\bar{c}_j^i\>\}_{j \in \bbZ_d}$ are the corresponding Schmidt bases. Then the target state $|\Psi\>$ can also be expressed as follows:
\begin{equation}
|\Psi\>=\frac{1}{\sqrt{d}}\sum_{i\in \bbZ_d} |\tilde{a}_i\>|\tilde{B}_i\> 
=\frac{1}{\sqrt{d}}\sum_{i\in \bbZ_d} \sum_{j\in \bbZ_d} u_j^i|\tilde{a}_i\>  |\bar{b}_j^i\> |\bar{c}_j^i\>. 
\end{equation}

Suppose the first party performs the projective measurement in the basis $\{|\tilde{a}_i\>\}_{i \in \bbZ_d}$ and obtains outcome $i$, then the reduced state of the remaining parties is $|\tilde{B}_i\>$. 
If the second party  performs the projective measurement in the Schmidt basis $\{|\bar{b}_j^i\>\}_{j \in \bbZ_d}$ for each outcome $i$, then the resulting test projector, labeled by the string $\bfm=(1,0)$, reads 
\begin{equation}
    P_{(1,0)} = \sum_{i\in \bbZ_d} \sum_{j \in \bbZ_d \mid u_j^i>0} |\tilde{a}_i\> \<\tilde{a}_i| \otimes |\bar{b}_j^i\> \<\bar{b}_j^i| \otimes |\bar{c}_j^i\> \<\bar{c}_j^i|.  \label{eq:P10}
\end{equation}
If instead the second party performs the projective measurement in the Fourier basis $\{|\gamma_j^i\>\}_{j \in \bbZ_d}$ with
\begin{equation}
    |\gamma_j^i\> = \sum_{l\in \bbZ_d} \frac{1}{\sqrt{d}} \omega^{jl} |\bar{b}_l^i\>
\end{equation}
for each outcome $i$, then  the resulting test projector, labeled by the string $\bfm=(1,1)$, reads
\begin{equation}
    P_{(1,1)} = \sum_{i\in \bbZ_d} \sum_{j\in \bbZ_d} |\tilde{a}_i\> \<\tilde{a}_i|  \otimes |\gamma_j^i\> \<\gamma_j^i| \otimes |\zeta_j^i\> \<\zeta_j^i|,  \label{eq:P11}
\end{equation}
where
\begin{equation}
    |\zeta^i_j\> = \sum_{l\in \bbZ_d} u_l^i \omega^{-jl} |\bar{c}_l^i\>.
\end{equation}
Equations~\eqref{eq:P00}, \eqref{eq:P01}, \eqref{eq:P10}, and \eqref{eq:P11} above exhaust all four test projectors featured in the SD protocol.

\subsection{Proof of \thref{thm:QSVSDgap}} \label{sec:thmQSVSDgapProof}

\begin{proof}[Proof of \thref{thm:QSVSDgap}]
	We shall prove \thref{thm:QSVSDgap} by induction on the number $n$ of parties. For simplicity of description, we assume that the total Hilbert space $\caH$ can be expressed as a tensor power $\caH=\caH_d^{\otimes n}$, where $\caH_d$ has dimension $d$, but the basic idea of our proof can be applied to a more general setting. 
	
	Suppose $|\Psi\>$ has the following decomposition with respect to the first party and the remaining $n-1$ parties:
	\begin{equation}
		|\Psi\>=\sum_{i\in \bbZ_d}s_i|a_i\>|B_i\>=\sum_{i\in \bbZ_d}\frac{1}{\sqrt{d}}|\ta_i\>|\tB_i\>;
	\end{equation}
	here the first decomposition is a SD, and $\{|\ta_i\>\}_{i \in \bbZ_d}$ is a basis on $\caH_d$ that is unbiased with the Schmidt basis $\{|a_i\>\}_{i \in \bbZ_d}$. To simplify the notation, the verification operator $\Omega_\SD$ is abbreviated as $\Omega$. By definition $\Omega$ can be expressed as follows:
	\begin{equation}
		\Omega=2^{1-n}\sum_{\bfm\in \bbZ_2^{n-1}} P_{\bfm},
	\end{equation}	
	where $P_\bfm$ are test projectors featured in the SD protocol.

	If  $n=2$, then $|\Psi\>$ is a bipartite pure state and we have $\Omega=(P_0+P_1)/2$, where the two test projectors $P_0$ and $P_1$ are given in Eq.~(4). It is straightforward to verify that $P_1$ has rank $d$ and $\tr(P_0P_1)=1$ \cite{LiHZ19}. Let $\bP_m=P_m-|\Psi\>\<\Psi|$ for $m=0,1$; then we have $\bar{\Omega}=\Omega-|\Psi\>\<\Psi|=(\bP_0+\bP_1)/2$. In addition,   $\bP_0$ and $\bP_1$ are mutually orthogonal projectors, and $\bP_1$ has rank $d-1\geq 1$.  Therefore,	
	\begin{equation}
		\beta(\Omega)=\|\bar{\Omega}\|=\frac{1}{2}\|\bP_0+\bP_1\|= \frac{1}{2},\quad
		\nu(\Omega)= \frac{1}{2}, \label{eq:nuLBbipartite}
	\end{equation}
	which implies \eref{eq:QSVSDNUB}.    
    If $|\Psi\>$ is entangled, then both $\bP_0$ and  $\bP_1$ have rank at least 1, which implies that
	\begin{equation}
		\|p\bP_0+(1-p)\bP_1\|=\max\{p,1-p\}\geq \frac{1}{2} 
	\end{equation}
	for $0\leq p\leq 1$. 
	So the choice $p=1/2$ (which means $P_0$ and $P_1$ are chosen with uniform probability) is optimal. 
	
	Next, suppose the result $\nu(\Omega)\geq 2^{1-n}$ holds for all $n$ with $2\leq n\leq \tn$ for some positive integer $\tn\geq 2$ and let $n=\tn+1$. Then $P_\bfm$  can be expressed as follows:
	\begin{equation}
		P_\bfm=\begin{cases}
			\sum_{i\in \bbZ_d \mid  s_i>0} |a_i\>\<a_i|\otimes P_{i,\bfm} & m_1=0,\\[1ex]
			\sum_{i\in \bbZ_d} |\ta_i\>\<\ta_i|\otimes \tP_{i,\bfm}  &m_1=1,
		\end{cases}
	\end{equation}
	where $P_{i,\bfm}$ ($\tP_{i,\bfm}$) are test projectors featured in the SD protocol applied to $|B_i\>$ ($|\tB_i\>$). Let 
    \begin{equation}
  	\begin{aligned}
	\!	\Omega_i&=2^{2-n}\sum_{\bfm\in \bbZ_2^{n-1} \mid m_1=0} P_{i,\bfm}\;\;   \forall \, i\in \bbZ_d,\, s_i>0,\\
      \!  \tOmega_i 
        &=2^{2-n}\sum_{\bfm\in \bbZ_2^{n-1} \mid m_1=1}\tP_{i,\bfm}\;\;  \forall \, i\in \bbZ_d;
	\end{aligned}      
    \end{equation}
then $\Omega_i$ ($\tOmega_i$) is a verification operator of $|B_i\>$ ($|\tB_i\>$). Let 
\begin{equation}
    \nu_k:=2^{1-k},\quad \beta_k:=1-\nu_k \quad \forall \, k=2,3,\ldots, n; \label{eq:nubetak}
\end{equation}
then, by induction hypothesis, we can deduce that 
\begin{equation}
    \begin{aligned}
    \Omega_i &\leq \nu_{n-1}|B_i\>\<B_i|+\beta_{n-1}, \\
    \tOmega_i &\leq \nu_{n-1}|\tB_i\>\<\tB_i|+\beta_{n-1}. 
    \end{aligned}
\end{equation}
Now, if $|\Psi\>$ is regarded as a bipartite pure state between the first party and the remaining $n-1$ parties, then the two test projectors featured in the SD protocol read [cf. Eq.~(4)]
\begin{equation}
\begin{aligned}
    \Pi_0 &= \sum_{i\in \bbZ_d \mid  s_i>0} |a_i\> \<a_i| \otimes |B_i\>\<B_i|,  \\
    \Pi_1 &= \sum_{i\in \bbZ_d} |\tilde{a}_i\> \<\tilde{a}_i| \otimes |\tilde{B}_i\>\<\tilde{B}_i|.
\end{aligned}
\end{equation}
In addition,  $(\Pi_0+\Pi_1)/2$ is a verification operator of $|\Psi\>\<\Psi|$ with spectral gap $1/2$ [cf.  \eref{eq:nuLBbipartite}], that is,
\begin{equation}
    |\Psi\>\<\Psi|\leq \frac{1}{2}(\Pi_0+\Pi_1)\leq \frac{1}{2}(|\Psi\>\<\Psi|+1). 
\end{equation}
	
	Combining the above results, we can deduce that
	\begin{align}
		\Omega&=\frac{1}{2}\sum_{i\in \bbZ_d \mid  s_i>0} |a_i\>\<a_i|\otimes \Omega_i +\frac{1}{2}\sum_{i\in \bbZ_d} |\ta_i\>\<\ta_i|\otimes \tOmega_i \nonumber\\
		&\leq 
		\frac{1}{2}\sum_{i\in \bbZ_d \mid  s_i>0}|a_i\>\<a_i|\otimes \left(\nu_{n-1}|B_i\>\<B_i|+\beta_{n-1}\right) \nonumber \\
         &\equad+\frac{1}{2}\sum_{i\in \bbZ_d} |\ta_i\>\<\ta_i|\otimes \left(\nu_{n-1}|\tB_i\>\<\tB_i|+\beta_{n-1}\right) \nonumber\\
		&\leq \frac{\nu_{n-1}}{2}(\Pi_0+\Pi_1)+\beta_{n-1} \nonumber \\
		&\leq \frac{\nu_{n-1}}{2}(|\Psi\>\<\Psi|+1)+\beta_{n-1} 
        \nonumber \\
        &= \nu_n |\Psi\>\<\Psi|+\beta_n,
	\end{align}
    where the parameters $\nu_k$ and $\beta_k$ for $k=n-1,n$ are defined in \eref{eq:nubetak}. 
	This equation implies that  $\nu(\Omega)\geq \nu_n= 2^{1-n}$, which in turn implies Eq.~(5)  and completes the proof of \thref{thm:QSVSDgap}. 
\end{proof}

\begin{figure}[b]
	\centering
	\includegraphics[width=0.49\textwidth]{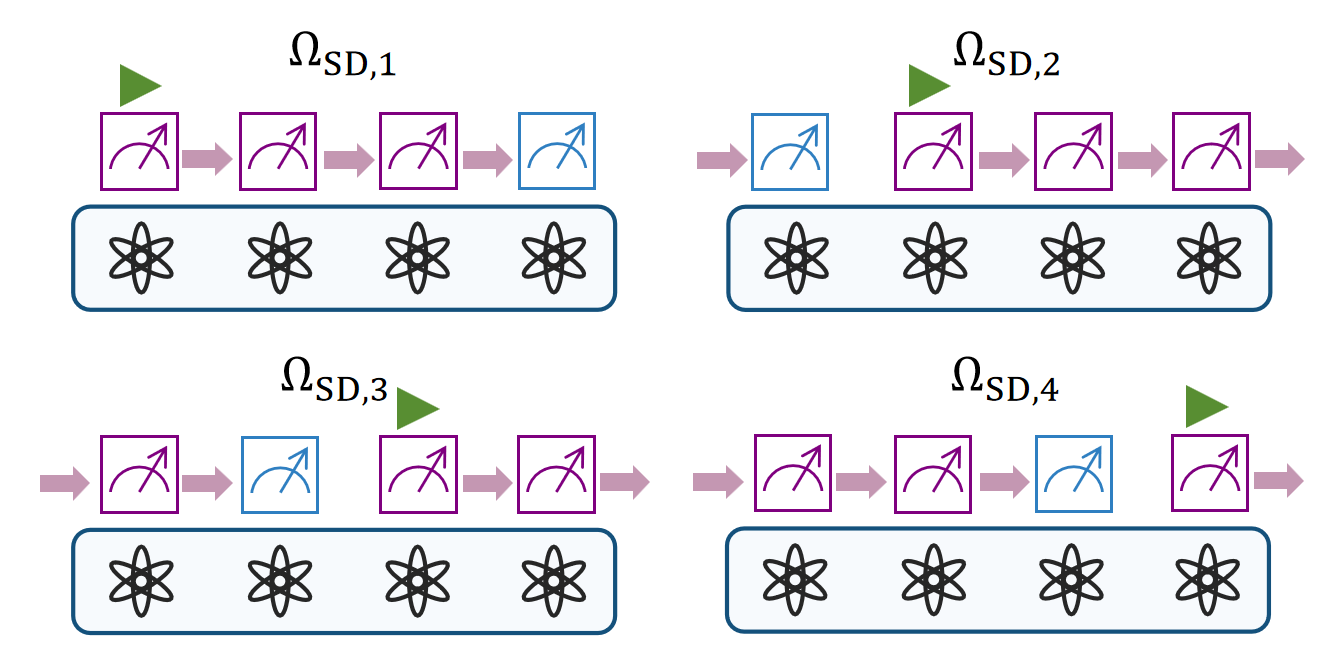}
	\caption{Schematic  of the CSD protocol as a probabilistic mixture of $n$ SD protocols tied to $n$ different orders of SD.  The first party is chosen randomly, which determines the order of SD and the corresponding SD protocol according to a cyclic sequence.}
	\label{fig:CSDframework}
\end{figure}

\subsection{Variants of the SD protocol}

The SD protocol presented in \sref{sec:SDpp} is based on a given order of successive SD, which determines the order of adaptive local projective measurements. Any permutation of the $n$ parties yields an equally good order. Moreover, the spectral gap can usually be increased by averaging over different orders. For example, here we propose the \emph{cyclic SD} (CSD) protocol, which is the average over  $n$ orders that are related by a cyclic permutation as illustrated in \fref{fig:CSDframework}. The resulting verification operator can be expressed as 
 $\Omega_{\CSD} = \sum_{k=1}^n \Omega_{\SD, k} / n$, where $\Omega_{\SD, k}$ denotes the verification operator 
 for which party $k$ starts the projective measurement. 
The spectral gap $\nu(\Omega_{\SD, k})$ averaged over Haar-random pure states is independent of $k$, so the CSD protocol can achieve a larger average spectral gap, as illustrated in \fref{fig:HaarSD}. When $d=4,n\geq 4$ or $d=5, n \geq 3$, the average spectral gap is even larger than $1/2$, so the performance of adaptive local projective measurements is quite close to the performance of entangling measurements for Haar-random pure states, which is quite surprising. Additional numerical results on Dicke states are shown in \fref{fig:Dicke_CSD} in \aref{app:CSDprotocol}. 

Incidentally, many other variants of the SD protocol are conceivable by considering more bases unbiased with the Schmidt basis  or more permutations. However, such protocols entail more measurement settings, and their advantages over the CSD protocol are limited,  given that the average spectral gap achieved by the CSD protocol is already quite large.

\section{MUB protocol}

\subsection{Protocol and performance}

\begin{figure}
	\centering
	\includegraphics[width=0.45\textwidth]{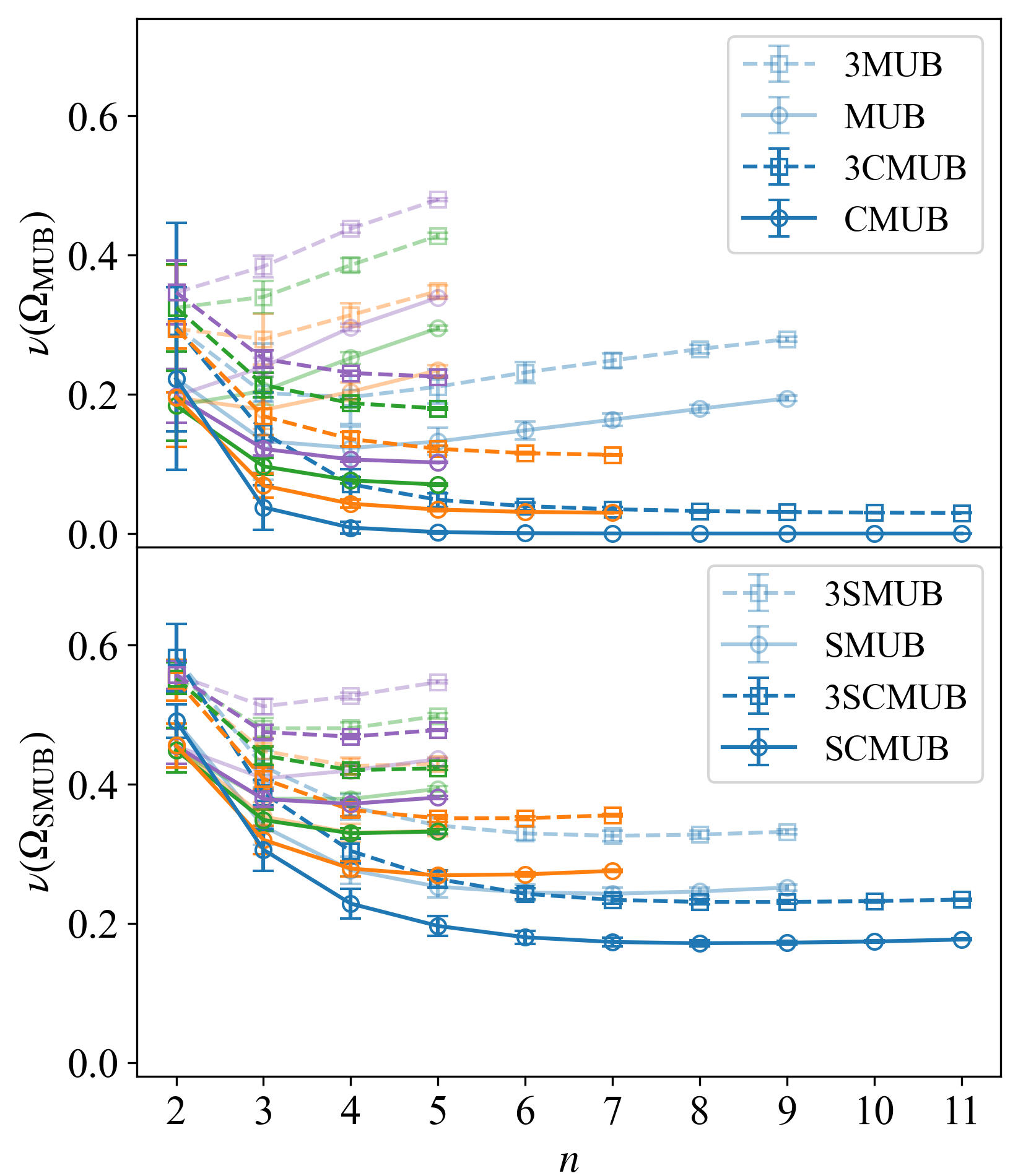}
	\caption{Average spectral gaps 
		achieved by eight variants of the MUB protocol  for $n$-qudit Haar-random pure states with $d=2,3,4,5$. The color encoding of the local dimension $d$ follows the lower plot in \fref{fig:HaarSD} (blue, orange, green, and purple mean $d=2,3,4,5$, respectively).}
	\label{fig:HaarMUB}
\end{figure}

In this section, we introduce a simpler protocol based on MUB to avoid successive SD. For concreteness, MUB can be constructed from the eigenbases of the two \emph{generalized Pauli operators}
$Z = \sum_{i\in \bbZ_d} \omega^i |i\>\<i|$ and   $X = \sum_{i\in \bbZ_d} {|i+1\>\<i|}$, where $\omega = \rme^{2\pi \rmi /d}$, although this choice is not essential. Note that the eigenbasis of $Z$ coincides with the computational basis. In the MUB protocol, each party among the first $n-1$ parties randomly chooses one basis from the MUB and performs the corresponding projective measurement, then party $n$ performs the  projective measurement onto  the conditional reduced state of the target state $|\Psi\>$ depending on the measurement choices and outcomes of the first $n-1$ parties. Like the SD protocol, the MUB protocol consists of $2^{n-1}$ tests. Each test can be labeled by a string $\bfm$ in $\bbZ_2^{n-1}$, where $m_k=0, 1$ indicate $Z, X$ measurements, respectively, for the $k$-th party; the resulting test projector is denoted by $Q_\bfm$ henceforth (see \sref{sec:TestProjMUB} for more details). If  $Q_\bfm$ is performed with probability $q_\bfm$, then the overall verification operator reads $\Omega_{\MUB} = \sum_{\bfm\in \bbZ_2^{n-1}} q_\bfm Q_\bfm$.  
Although the MUB protocol is substantially simpler, it can achieve similar performance to  the SD protocol, as illustrated in \fref{fig:HaarMUB} (see also \fref{fig:HaarSD} and \aref{app:compareProtocols}). The deep reason behind this similarity merits further exploration. It is also
desirable to gain an analytical understanding of the average performance of the MUB protocol.

\subsection{Variants of the MUB protocol}

\begin{figure}[bt]
	\centering
	\includegraphics[width=0.4\textwidth]{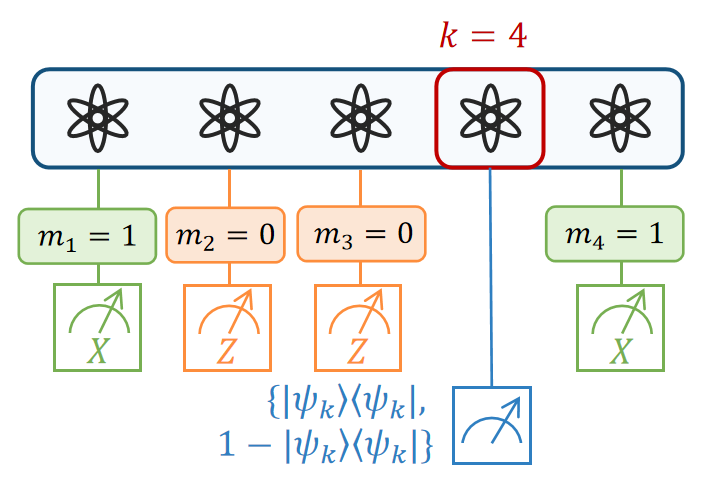}
	\caption{Schematic of the SMUB protocol as a probabilistic mixture of $n$ MUB protocols. Each MUB protocol featured in the SMUB protocol is specified by the choice of the last party (here $k=4$), the party that performs the projective measurement at last conditioned on the measurement choices [specified by the string $\bfm=(1,0,0,1)$] and outcomes of the other $n-1$ parties. The projective measurement of the last party is determined by the conditional reduced state $|\psi_k\>=|\psi_{k,\bfo, \bfm}\>$ of the target state. }
	\label{fig:SMUBframework}
\end{figure}

As an alternative, the first $n-1$ parties can perform $Z$ ($X$) measurements simultaneously. In this way, we can construct a \emph{correlated MUB} (CMUB) protocol, which can reduce the number of distinct tests from $2^{n-1}$ to 2. Also, instead of party $n$, party $k$ for $k=1,2,\ldots, n-1$ can  serve as the last party (who performs the projective measurement at last), and the resulting verification operator is denoted by $\Omega_{\MUB,k}$ henceforth. On this basis we can construct  a \emph{symmetrized MUB} (SMUB) protocol by randomizing the choice of the last party  as illustrated  in \fref{fig:SMUBframework}, and the resulting verification operator  can be expressed as $\Omega_{\SMUB}=\sum_{k=1}^n \Omega_{\MUB,k}/n$. Symmetrized CMUB protocol (SCMUB) can be constructed in a similarly way. In addition, each of the first $n-1$ parties can perform three or more projective measurements that are mutually unbiased. For example, the respective eigenbases of  $X$, $Z$, and  $Y=\rmi XZ$ are mutually unbiased, from which we can construct the 3MUB protocol. Likewise, 3CMUB, symmetrized 3MUB (3SMUB), and symmetrized 3CMUB (3SCMUB) protocols can be constructed.

The performance of the above variants for Haar-random pure states is  shown in \fref{fig:HaarMUB}. 
Surprisingly, the CMUB protocol (with only two distinct tests) can verify generic Haar-random pure states; moreover, its  performance is comparable to the MUB protocol when $d\geq 3$. 
Symmetrization and more basis choices can usually increase the spectral gap but do not change the overall performance significantly (when $d=2$, the improvements of 3CMUB and SCMUB over CMUB are exceptional). 
Additional results on Dicke states are shown in \asref{app:MUBCMUBprotocols} and \ref{app:SMUBSCMUBprotocols}, which include \fsref{fig:DickeMUB2B}-\ref{fig:Dicke3SMUBd2} and \tref{tab:DickeSDP}. Comparisons of various verification protocols are summarized in \asref{app:compareProtocols} and \ref{app:beyondMUB}, which include \fsref{fig:hboxCompare}-\ref{fig:HaarPlantonic} and \tsref{tab:compareHaardn} and \ref{tab:compareDickedn}.

\subsection{Test projectors and verification operators featured in the MUB and CMUB protocols}
\label{sec:TestProjMUB}

For concreteness, here we provide explicit expressions for the test projectors and verification operators of the MUB and CMUB protocols. See
\aref{app:SMUBSCMUBprotocols} for the counterparts of the SMUB and SCMUB protocols.

Recall that the MUB protocol consists of $2^{n-1}$ projective tests, and each test  is labeled by a string $\bfm$ in $\bbZ_2^{n-1}$, where $m_k=0$  means the measurement in the computational basis (eigenbasis of $Z$) for party $k$, while  $m_k=1$ means the measurement in a basis mutually unbiased with the computational basis, say, the eigenbasis of  $X$. The corresponding test projector $Q_\bfm$ can be expressed as follows:
\begin{align}
    Q_\bfm
    = \sum_{\bfo \in \bbZ_d^{n-1}} &\Pi_{o_1, m_1} \otimes \cdots \otimes \Pi_{o_{n-1}, m_{n-1}} \nonumber \\
    &\equad \otimes |\psi_{n,\bfo, \bfm}\>\<\psi_{n,\bfo, \bfm}|, \label{eq:QmGPauli}
\end{align}
where $\Pi_{o_k, m_k}$ for each $k=1,2,\ldots, n-1$ denotes the projector onto the $o_k$-th basis state associated with the computational basis (when $m_k=0$) or MUB (when $m_k=1$), and $|\psi_{n,\bfo, \bfm}\>$ is the conditional reduced state of the target state $|\Psi\>$ for party $n$ given the measurement setting $\bfm$ and outcome $\bfo$.
Suppose the measurement setting $\bfm$ is chosen with probability $q_\bfm$ with $\sum_{\bfm\in \bbZ_2^{n-1}} q_\bfm = 1$; then the overall verification operator reads
\begin{equation}
    \Omega_{\MUB} = \sum_{\bfm\in \bbZ_2^{n-1}} q_\bfm Q_\bfm.
\end{equation}
For the uniform  distribution, that is, $q_\bfm = 2^{1-n}$ for all $\bfm \in \bbZ_2^{n-1}$, we have $\Omega_{\MUB}=2^{1-n}\sum_{\bfm\in \bbZ_2^{n-1}}  Q_\bfm$. By contrast, the CMUB protocol consists of only two tests labeled by the strings $\bfm=0^{n-1}$ and $\bfm=1^{n-1}$, respectively.

Recall that the respective eigenbases of  $X$, $Z$, and  $Y=\rmi XZ$ are mutually unbiased, from which we can construct the 3MUB protocol and 3CMUB protocol (when the measurement bases are correlated).  If in addition $d$ is an odd prime, then the respective eigenbases of $X, XZ, XZ^2, \dots, XZ^{d-1}, Z$ form a complete set of MUB, from which we can construct complete MUB (cMUB) and complete CMUB (cCMUB) protocols. The test projectors and verification operators of the 3MUB, 3CMUB,  cMUB, and cCMUB protocols  can be constructed in a similar way.

\section{Comparison with the HPS protocol}

Recently, HPS proposed a protocol that can certify almost all $n$-qubit pure  states using single-qubit measurements \cite{huang2025certifying}. In general, it is impossible  to formulate this protocol in the framework of standard QSV. To make fair comparison here we focus on a particular variant based on $n$ tests for which this formulation is possible \cite{li2025new}: all the parties except for one party (chosen randomly) perform Pauli $Z$ measurements, then the last party performs the projective measurement onto the conditional reduced state. Generalization to systems with higher local dimensions is straightforward, but a rigorous performance guarantee is not yet available.

\begin{figure}
	\centering
	\includegraphics[width=0.45\textwidth]{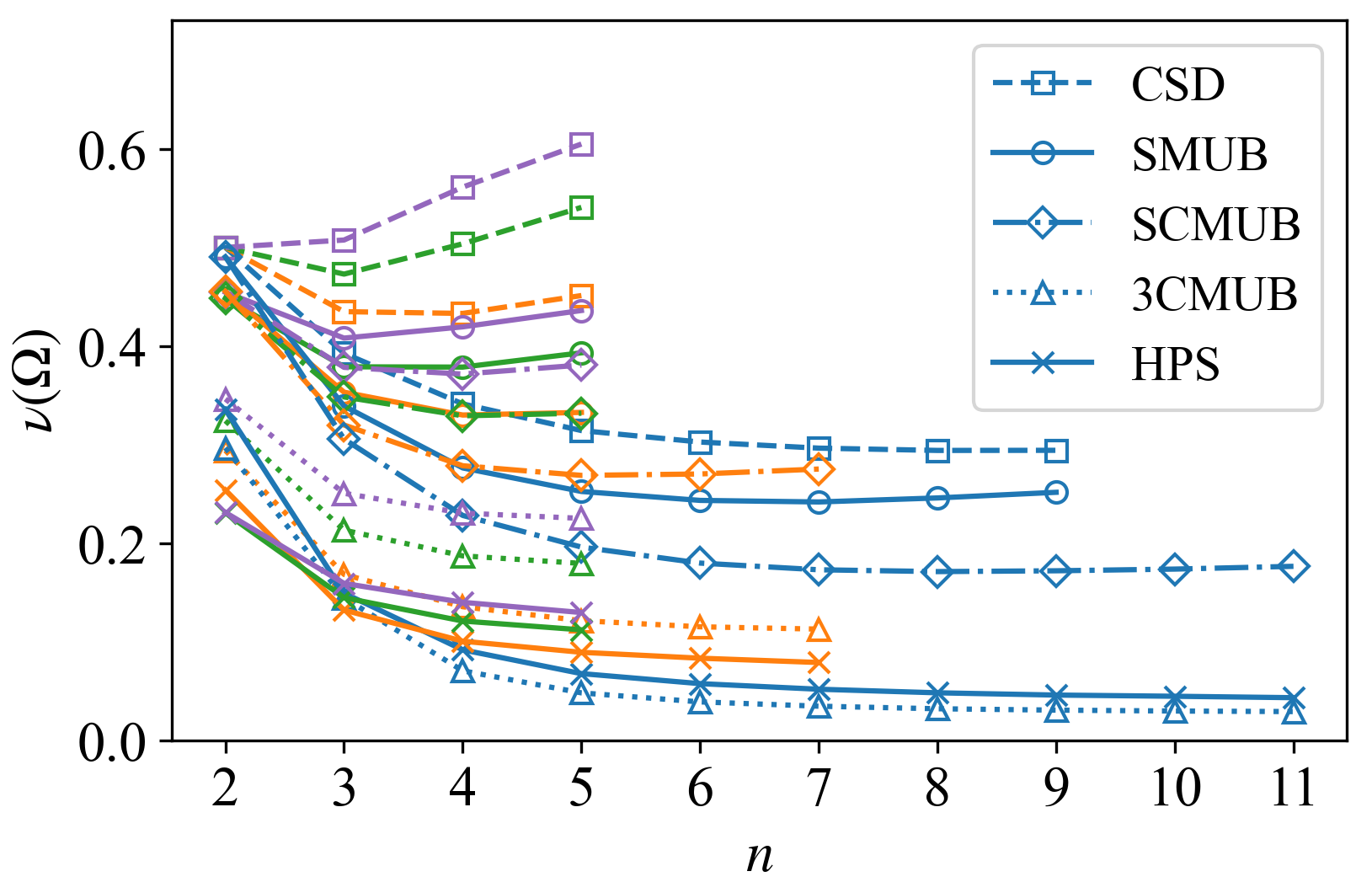}
	\caption{
		Average spectral gaps achieved by the CSD, SMUB, SCMUB, and 3CMUB  protocols in comparison with the HPS protocol \cite{huang2025certifying} for $n$-qudit Haar-random pure states with $d=2,3,4,5$.  The color encoding follows \fref{fig:HaarMUB}. }
	\label{fig:GapSDMUBHPS}
\end{figure}

\Fref{fig:GapSDMUBHPS} compares the performance of the CSD, SMUB, SCMUB,  3CMUB, and HPS protocols for Haar-random pure states (see also \fref{fig:GapRatioHPS} in \aref{app:compareHPS}). The first three protocols can achieve much larger average spectral gaps, which are lower bounded  by a universal constant according to  numerical calculations, while the asymptotic behavior (for large $n$) of the HPS protocol is not clear. When  $d=2$ and $n=9$ for example,  the average spectral gaps achieved by the CSD and SMUB protocols are more than five times as large as the counterpart for the HPS protocol. The 3CMUB protocol can  achieve average spectral gaps comparable to those of the HPS protocol but requires only three distinct tests.
Further comparisons are shown in \tref{tab:compareHuang} in \aref{app:compareHPS}.

\section{Universal QSV in the adversarial scenario}

\begin{figure}
	\centering
    \vspace{1ex}
	\includegraphics[width=0.45\textwidth]{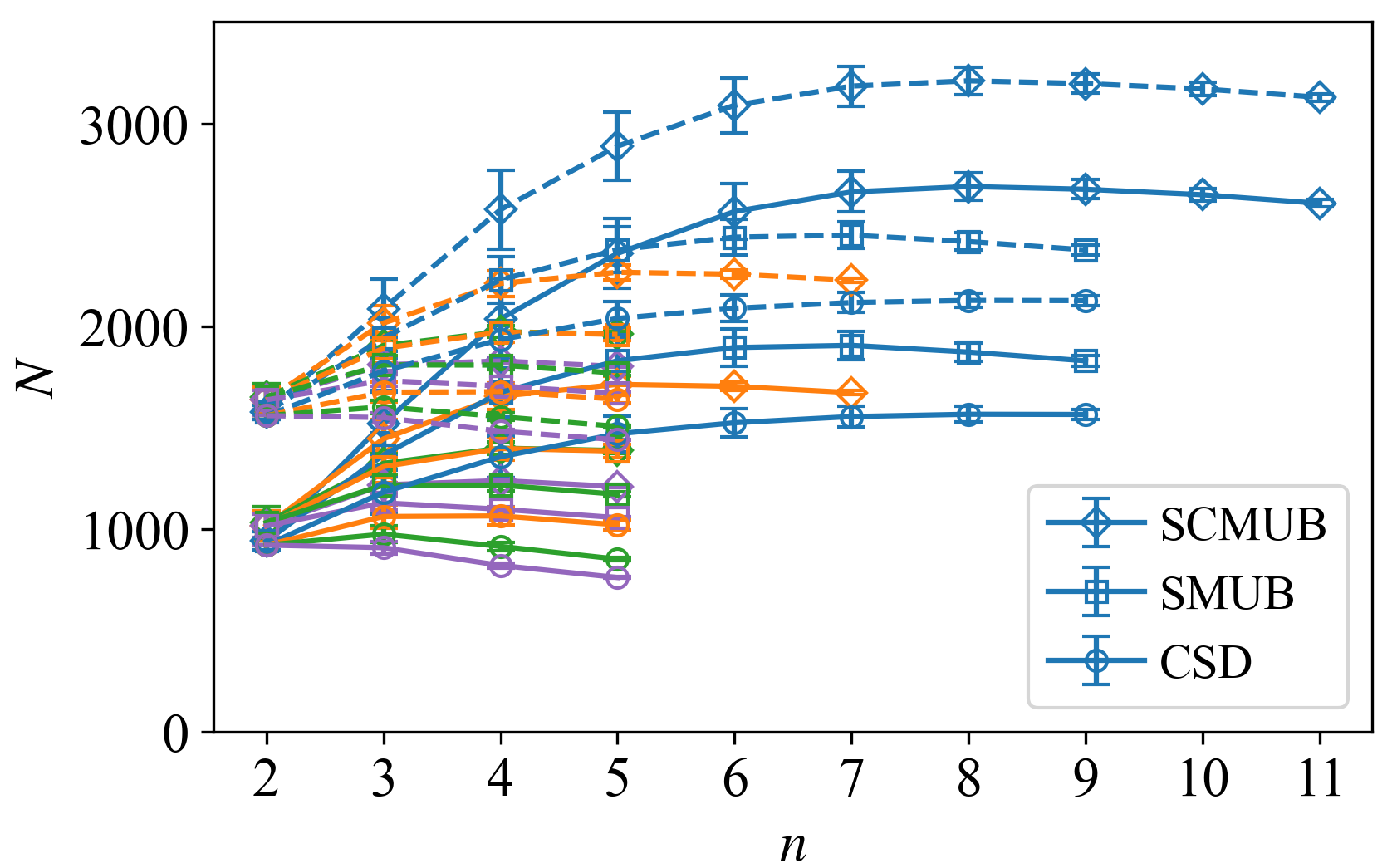}
	\caption{Average sample costs required by the CSD, SMUB, and SCMUB protocols for  verifying $n$-qudit Haar-random pure states with $d=2,3,4,5$ within precision $\delta=\varepsilon=0.01$ in the nonadversarial (solid lines) and adversarial (dashed lines) scenarios. The color encoding follows \fref{fig:HaarMUB}. }
	\label{fig:HaarAdv}
\end{figure}

Now, the preparation device may be controlled by a  malicious adversary, and QSV is much more challenging. Fortunately, any verification protocol for the nonadversarial scenario can be extended to the adversarial scenario with a small sample overhead thanks to Theorem~6 in \rcite{Zhu2019Efficient}. Suppose $\Omega$ is a verification operator for $|\Psi\>$ with spectral gap $\nu$; let $p=\nu/\rme$ and $\Omega_p=p+(1-p)\Omega$, where $\rme$ is the base of the natural logarithm. Then $|\Psi\>$ can be verified efficiently in the adversarial scenario by virtue of $\Omega_p$ and random permutations, and the number of tests required to achieve infidelity $\varepsilon$ and significance level $\delta$ satisfies~\cite{Zhu2019Efficient}
\begin{equation}
    N \le \frac{\ln[(1-\varepsilon)^{-1}\delta^{-1}]}{(1-\nu + \rme^{-1} \nu^2)\nu \varepsilon}.
\end{equation}
If $\Omega$ corresponds  to the SD protocol, then we have $\nu(\Omega)\geq 2^{1-n}$ by \thref{thm:QSVSDgap}, so  $N\leq 2^n \varepsilon^{-1}\ln (\delta^{-1})$ whenever $\varepsilon,\delta\leq 1/5$. Numerical simulations further show that
the CSD, SMUB, and SCMUB protocols can   verify Haar-random pure states with constant sample costs in both the nonadversarial and adversarial scenarios, as illustrated in \fref{fig:HaarAdv}.

\section{Summary}

We proposed a universal protocol, the SD protocol, for verifying arbitrary multipartite pure quantum states using adaptive local projective measurements. Its performance  is guaranteed by a rigorous upper bound on the sample complexity,  independent of the local dimensions. Numerical calculations further show that this protocol can verify Haar-random pure states with a constant sample cost even in the adversarial scenario. As alternatives, we also introduced several simpler  variants based on MUB, which can achieve  similar performance. Numerical calculations indicate that a generic Haar-random pure state can be verified efficiently even if we can only access a constant number of distinct tests based on local projective measurements. Our work offers valuable insights on the power of local projective measurements in QSV and useful tools for verifying multipartite pure states, which may find applications in many tasks in quantum information processing, such as quantum benchmarking, quantum computation, quantum simulation, and quantum machine learning.

Our work also opens several promising potential research directions. Specifically, establishing informative analytical lower bounds for the average spectral gaps achieved by variants of the SD and MUB protocols is a critical next step. Additionally, exploring verification protocols beyond MUB and investigating the connection between quantum incompatibility and sample complexity present significant opportunities for future research.
Furthermore, determining the fundamental efficiency limit of local projective measurements remains a key unresolved challenge.

\bigskip

\textbf{Note added.} Upon completion of this work, we 
noted \rcite{gupta2025}, which proved that any $n$-qubit pure state can be certified using $\caO(n)$ samples.

\bigskip

The data and codes for this study are available at https://github.com/WinkinTing/UQSV.

\acknowledgments
We thank Zihao Li,  Ye-Chao Liu, Xiao-Dong Yu, and Jiangwei Shang for valuable discussions. This work is supported by Shanghai Science and Technology Innovation Action Plan (Grant No.~24LZ1400200), Shanghai Municipal Science and Technology Major Project (Grant No.~2019SHZDZX01), and the National Key Research and Development Program of China (Grant No.~2022YFA1404204).

\bibliographystyle{quantum}
\bibliography{RSV_ref.bib}

\begin{thebibliography}{10}

\bibitem{Horodecki2009}
Ryszard Horodecki, Pawe\l{} Horodecki, Micha\l{} Horodecki, and Karol
  Horodecki.
\newblock ``Quantum entanglement''.
\newblock \href{https://dx.doi.org/10.1103/RevModPhys.81.865}{Rev. Mod. Phys.
  {\bf 81}, 865--942}~(2009).

\bibitem{GUHNE2009}
Otfried Gühne and Géza Tóth.
\newblock ``Entanglement detection''.
\newblock
  \href{https://dx.doi.org/https://doi.org/10.1016/j.physrep.2009.02.004}{Phys.
  Rep. {\bf 474}, 1--75}~(2009).

\bibitem{Walter2016}
Michael Walter, David Gross, and Jens Eisert.
\newblock ``Multipartite entanglement''.
\newblock
  \href{https://dx.doi.org/https://doi.org/10.1002/9783527805785.ch14}{Chapter~14,
  pages 293--330}.
\newblock John Wiley \& Sons. ~(2016).

\bibitem{CarrEKK21}
Jose Carrasco, Andreas Elben, Christian Kokail, Barbara Kraus, and Peter
  Zoller.
\newblock ``Theoretical and experimental perspectives of quantum
  verification''.
\newblock \href{https://dx.doi.org/10.1103/PRXQuantum.2.010102}{PRX Quantum
  {\bf 2}, 010102}~(2021).

\bibitem{MorrSGD22}
Joshua Morris, Valeria Saggio, Aleksandra Gočanin, and Borivoje Dakić.
\newblock ``Quantum verification and estimation with few copies''.
\newblock \href{https://dx.doi.org/https://doi.org/10.1002/qute.202100118}{Adv.
  Quantum Technol. {\bf 5}, 2100118}~(2022).

\bibitem{YuSG22}
Xiao-Dong Yu, Jiangwei Shang, and Otfried Gühne.
\newblock ``Statistical methods for quantum state verification and fidelity
  estimation''.
\newblock \href{https://dx.doi.org/https://doi.org/10.1002/qute.202100126}{Adv.
  Quantum Technol. {\bf 5}, 2100126}~(2022).

\bibitem{HayaMT06}
Masahito Hayashi, Keiji Matsumoto, and Yoshiyuki Tsuda.
\newblock ``A study of {LOCC}-detection of a maximally entangled state using
  hypothesis testing''.
\newblock \href{https://dx.doi.org/10.1088/0305-4470/39/46/013}{J. Phys. A:
  Math. Gen. {\bf 39}, 14427}~(2006).

\bibitem{AoliGKE15}
Leandro Aolita, Christian Gogolin, Martin Kliesch, and Jens Eisert.
\newblock ``Reliable quantum certification of photonic state preparations''.
\newblock \href{https://dx.doi.org/10.1038/ncomms9498}{Nat. Commun. {\bf 6},
  8498}~(2015).

\bibitem{HangKSE17}
Dominik Hangleiter, Martin Kliesch, Martin Schwarz, and Jens Eisert.
\newblock ``Direct certification of a class of quantum simulations''.
\newblock \href{https://dx.doi.org/10.1088/2058-9565/2/1/015004}{Quantum Sci.
  Technol. {\bf 2}, 015004}~(2017).

\bibitem{TakeM18}
Yuki Takeuchi and Tomoyuki Morimae.
\newblock ``Verification of many-qubit states''.
\newblock \href{https://dx.doi.org/10.1103/PhysRevX.8.021060}{Phys. Rev. X {\bf
  8}, 021060}~(2018).

\bibitem{PallLM18}
Sam Pallister, Noah Linden, and Ashley Montanaro.
\newblock ``Optimal verification of entangled states with local measurements''.
\newblock \href{https://dx.doi.org/10.1103/PhysRevLett.120.170502}{Phys. Rev.
  Lett. {\bf 120}, 170502}~(2018).

\bibitem{Zhu2019Efficient}
Huangjun Zhu and Masahito Hayashi.
\newblock ``Efficient verification of pure quantum states in the adversarial
  scenario''.
\newblock \href{https://dx.doi.org/10.1103/PhysRevLett.123.260504}{Phys. Rev.
  Lett. {\bf 123}, 260504}~(2019).

\bibitem{Zhu2019General}
Huangjun Zhu and Masahito Hayashi.
\newblock ``General framework for verifying pure quantum states in the
  adversarial scenario''.
\newblock \href{https://dx.doi.org/10.1103/PhysRevA.100.062335}{Phys. Rev. A
  {\bf 100}, 062335}~(2019).

\bibitem{WuBGL21}
Ya-Dong Wu, Ge~Bai, Giulio Chiribella, and Nana Liu.
\newblock ``Efficient verification of continuous-variable quantum states and
  devices without assuming identical and independent operations''.
\newblock \href{https://dx.doi.org/10.1103/PhysRevLett.126.240503}{Phys. Rev.
  Lett. {\bf 126}, 240503}~(2021).

\bibitem{LiuSHZ21}
Ye-Chao Liu, Jiangwei Shang, Rui Han, and Xiangdong Zhang.
\newblock ``Universally optimal verification of entangled states with
  nondemolition measurements''.
\newblock \href{https://dx.doi.org/10.1103/PhysRevLett.126.090504}{Phys. Rev.
  Lett. {\bf 126}, 090504}~(2021).

\bibitem{GocaSD22}
Aleksandra Go\ifmmode~\check{c}\else \v{c}\fi{}anin, Ivan \ifmmode
  \check{S}\else \v{S}\fi{}upi\ifmmode~\acute{c}\else \'{c}\fi{}, and Borivoje
  Daki\ifmmode~\acute{c}\else \'{c}\fi{}.
\newblock ``Sample-efficient device-independent quantum state verification and
  certification''.
\newblock \href{https://dx.doi.org/10.1103/PRXQuantum.3.010317}{PRX Quantum
  {\bf 3}, 010317}~(2022).

\bibitem{Haya09G}
Masahito Hayashi.
\newblock ``Group theoretical study of {LOCC}-detection of maximally entangled
  states using hypothesis testing''.
\newblock \href{https://dx.doi.org/10.1088/1367-2630/11/4/043028}{New J. Phys.
  {\bf 11}, 043028}~(2009).

\bibitem{ZhuH19O}
Huangjun Zhu and Masahito Hayashi.
\newblock ``Optimal verification and fidelity estimation of maximally entangled
  states''.
\newblock \href{https://dx.doi.org/10.1103/PhysRevA.99.052346}{Phys. Rev. A
  {\bf 99}, 052346}~(2019).

\bibitem{WangH19}
Kun Wang and Masahito Hayashi.
\newblock ``Optimal verification of two-qubit pure states''.
\newblock \href{https://dx.doi.org/10.1103/PhysRevA.100.032315}{Phys. Rev. A
  {\bf 100}, 032315}~(2019).

\bibitem{LiHZ19}
Zihao Li, Yun-Guang Han, and Huangjun Zhu.
\newblock ``Efficient verification of bipartite pure states''.
\newblock \href{https://dx.doi.org/10.1103/PhysRevA.100.032316}{Phys. Rev. A
  {\bf 100}, 032316}~(2019).

\bibitem{YuSG19}
Xiao-Dong Yu, Jiangwei Shang, and Otfried G\"uhne.
\newblock ``{Optimal verification of general bipartite pure states}''.
\newblock \href{https://dx.doi.org/10.1038/s41534-019-0226-z}{npj Quantum Inf.
  {\bf 5}, 112}~(2019).

\bibitem{HayaM15}
Masahito Hayashi and Tomoyuki Morimae.
\newblock ``Verifiable measurement-only blind quantum computing with stabilizer
  testing''.
\newblock \href{https://dx.doi.org/10.1103/PhysRevLett.115.220502}{Phys. Rev.
  Lett. {\bf 115}, 220502}~(2015).

\bibitem{FujiH17}
Keisuke Fujii and Masahito Hayashi.
\newblock ``Verifiable fault tolerance in measurement-based quantum
  computation''.
\newblock \href{https://dx.doi.org/10.1103/PhysRevA.96.030301}{Phys. Rev. A
  {\bf 96}, 030301(R)}~(2017).

\bibitem{HayaH18}
Masahito Hayashi and Michal Hajdu\ifmmode~\check{s}\else \v{s}\fi{}ek.
\newblock ``Self-guaranteed measurement-based quantum computation''.
\newblock \href{https://dx.doi.org/10.1103/PhysRevA.97.052308}{Phys. Rev. A
  {\bf 97}, 052308}~(2018).

\bibitem{Markham2018ASP}
Damian Markham and Alexandra Krause.
\newblock ``A simple protocol for certifying graph states and applications in
  quantum networks''.
\newblock \href{https://dx.doi.org/10.3390/cryptography4010003}{Cryptogr. {\bf
  4}, 3}~(2020).

\bibitem{Li2020GHZ}
Zihao Li, Yun-Guang Han, and Huangjun Zhu.
\newblock ``Optimal verification of {Greenberger-Horne-Zeilinger} states''.
\newblock \href{https://dx.doi.org/10.1103/PhysRevApplied.13.054002}{Phys. Rev.
  Appl. {\bf 13}, 054002}~(2020).

\bibitem{ZhuH19E}
Huangjun Zhu and Masahito Hayashi.
\newblock ``Efficient verification of hypergraph states''.
\newblock \href{https://dx.doi.org/10.1103/PhysRevApplied.12.054047}{Phys. Rev.
  Appl. {\bf 12}, 054047}~(2019).

\bibitem{LiuYSZ19}
Ye-Chao Liu, Xiao-Dong Yu, Jiangwei Shang, Huangjun Zhu, and Xiangdong Zhang.
\newblock ``Efficient verification of {Dicke} states''.
\newblock \href{https://dx.doi.org/10.1103/PhysRevApplied.12.044020}{Phys. Rev.
  Appl. {\bf 12}, 044020}~(2019).

\bibitem{Li2021Dicke}
Zihao Li, Yun-Guang Han, Hao-Feng Sun, Jiangwei Shang, and Huangjun Zhu.
\newblock ``Verification of phased {Dicke} states''.
\newblock \href{https://dx.doi.org/10.1103/PhysRevA.103.022601}{Phys. Rev. A
  {\bf 103}, 022601}~(2021).

\bibitem{LiYT2023AKLT}
Tianyi Chen, Yunting Li, and Huangjun Zhu.
\newblock ``Efficient verification of {Affleck-Kennedy-Lieb-Tasaki} states''.
\newblock \href{https://dx.doi.org/10.1103/PhysRevA.107.022616}{Phys. Rev. A
  {\bf 107}, 022616}~(2023).

\bibitem{Zhu2024FFH}
Huangjun Zhu, Yunting Li, and Tianyi Chen.
\newblock ``Efficient verification of ground states of frustration-free
  {H}amiltonians''.
\newblock \href{https://dx.doi.org/10.22331/q-2024-01-10-1221}{{Quantum} {\bf
  8}, 1221}~(2024).

\bibitem{ZhanZCP20}
Wen-Hao Zhang, Chao Zhang, Zhe Chen, Xing-Xiang Peng, Xiao-Ye Xu, Peng Yin,
  Shang Yu, Xiang-Jun Ye, Yong-Jian Han, Jin-Shi Xu, Geng Chen, Chuan-Feng Li,
  and Guang-Can Guo.
\newblock ``Experimental optimal verification of entangled states using local
  measurements''.
\newblock \href{https://dx.doi.org/10.1103/PhysRevLett.125.030506}{Phys. Rev.
  Lett. {\bf 125}, 030506}~(2020).

\bibitem{ZhanLYP20}
Wen-Hao Zhang, Xiao Liu, Peng Yin, Xing-Xiang Peng, Gong-Chu Li, Xiao-Ye Xu,
  Shang Yu, Zhi-Bo Hou, Yong-Jian Han, Jin-Shi Xu, Zong-Quan Zhou, Geng Chen,
  Chuan-Feng Li, and Guang-Can Guo.
\newblock ``Classical communication enhanced quantum state verification''.
\newblock \href{https://dx.doi.org/10.1038/s41534-020-00328-4}{npj Quantum Inf.
  {\bf 6}, 103}~(2020).

\bibitem{LuXCC20}
Liangliang Lu, Lijun Xia, Zhiyu Chen, Leizhen Chen, Tonghua Yu, Tao Tao,
  Wenchao Ma, Ying Pan, Xinlun Cai, Yanqing Lu, Shining Zhu, and Xiao-Song Ma.
\newblock ``Three-dimensional entanglement on a silicon chip''.
\newblock \href{https://dx.doi.org/10.1038/s41534-020-0260-x}{npj Quantum Inf.
  {\bf 6}, 30}~(2020).

\bibitem{JianWQC20}
Xinhe Jiang, Kun Wang, Kaiyi Qian, Zhaozhong Chen, Zhiyu Chen, Liangliang Lu,
  Lijun Xia, Fangmin Song, Shining Zhu, and Xiao-Song Ma.
\newblock ``Towards the standardization of quantum state verification using
  optimal strategies''.
\newblock \href{https://dx.doi.org/10.1038/s41534-020-00317-7}{npj Quantum Inf.
  {\bf 6}, 90}~(2020).

\bibitem{huang2025certifying}
Hsin-Yuan Huang, John Preskill, and Mehdi Soleimanifar.
\newblock ``Certifying almost all quantum states with few single-qubit
  measurements''.
\newblock \href{https://dx.doi.org/10.1038/s41567-025-03025-1}{Nat. Phys. {\bf
  21}, 1834--1841}~(2025).

\bibitem{Liu2023Homo}
Ye-Chao Liu, Yinfei Li, Jiangwei Shang, and Xiangdong Zhang.
\newblock ``Efficient verification of arbitrary entangled states with
  homogeneous local measurements''.
\newblock \href{https://dx.doi.org/10.1002/qute.202300083}{Adv. Quantum
  Technol. {\bf 6}, 2300083}~(2023).

\bibitem{NielC10book}
Michael~A. Nielsen and Isaac~L. Chuang.
\newblock ``Quantum computation and quantum information''.
\newblock \href{https://dx.doi.org/10.1017/CBO9780511976667}{Cambridge
  University Press}. Cambridge, UK~(2010).
\newblock 2nd edition.

\bibitem{Schwinger1960}
Julian Schwinger.
\newblock ``Unitary operator bases''.
\newblock \href{https://dx.doi.org/10.1073/pnas.46.4.570}{Proc. Natl. Acad.
  Sci. {\bf 46}, 570--579}~(1960).

\bibitem{Ivonovic1981}
Igor~D. Ivanovic.
\newblock ``Geometrical description of quantal state determination''.
\newblock \href{https://dx.doi.org/10.1088/0305-4470/14/12/019}{J. Phys. A:
  Math. Gen. {\bf 14}, 3241}~(1981).

\bibitem{WOOTTERS1989363}
William~K Wootters and Brian~D Fields.
\newblock ``Optimal state-determination by mutually unbiased measurements''.
\newblock
  \href{https://dx.doi.org/https://doi.org/10.1016/0003-4916(89)90322-9}{Ann.
  Phys. (N. Y.) {\bf 191}, 363--381}~(1989).

\bibitem{DURT2010}
Thomas Durt, Berthold-Georg Englert, Ingemar Bengtsson, and Karol
  \.{Z}yczkowski.
\newblock ``On mutually unbiased bases''.
\newblock \href{https://dx.doi.org/10.1142/S0219749910006502}{Int. J. Quantum
  Inf. {\bf 08}, 535--640}~(2010).

\bibitem{RausB01}
Robert Raussendorf and Hans~J. Briegel.
\newblock ``A one-way quantum computer''.
\newblock \href{https://dx.doi.org/10.1103/PhysRevLett.86.5188}{Phys. Rev.
  Lett. {\bf 86}, 5188--5191}~(2001).

\bibitem{BroaFK09}
Anne Broadbent, Joseph Fitzsimons, and Elham Kashefi.
\newblock ``Universal blind quantum computation''.
\newblock In 2009 50th Annual IEEE Symposium on Foundations of Computer
  Science.
\newblock \href{https://dx.doi.org/10.1109/FOCS.2009.36}{Pages 517--526}.
\newblock ~(2009).
\newblock  \href{http://arxiv.org/abs/0807.4154}{arXiv:0807.4154}.

\bibitem{Fitzsimons2017}
Joseph~F. Fitzsimons.
\newblock ``Private quantum computation: an introduction to blind quantum
  computing and related protocols''.
\newblock \href{https://dx.doi.org/10.1038/s41534-017-0025-3}{npj Quantum Inf.
  {\bf 3}, 23}~(2017).

\bibitem{LiZH23}
Zihao Li, Huangjun Zhu, and Masahito Hayashi.
\newblock ``Robust and efficient verification of graph states in blind
  measurement-based quantum computation''.
\newblock \href{https://dx.doi.org/10.1038/s41534-023-00783-9}{npj Quantum Inf.
  {\bf 9}, 115}~(2023).

\bibitem{vandenberghe1996semidefinite}
Lieven Vandenberghe and Stephen Boyd.
\newblock ``Semidefinite programming''.
\newblock \href{https://dx.doi.org/10.1137/1038003}{SIAM Rev. {\bf 38},
  49--95}~(1996).

\bibitem{Roy2007}
Aidan Roy and A.~J. Scott.
\newblock ``{Weighted complex projective 2-designs from bases: Optimal state
  determination by orthogonal measurements}''.
\newblock \href{https://dx.doi.org/10.1063/1.2748617}{J. Math. Phys. {\bf 48},
  072110}~(2007).

\bibitem{Adamson2010}
R.~B.~A. Adamson and A.~M. Steinberg.
\newblock ``Improving quantum state estimation with mutually unbiased bases''.
\newblock \href{https://dx.doi.org/10.1103/PhysRevLett.105.030406}{Phys. Rev.
  Lett. {\bf 105}, 030406}~(2010).

\bibitem{Zhu2014}
Huangjun Zhu.
\newblock ``Quantum state estimation with informationally overcomplete
  measurements''.
\newblock \href{https://dx.doi.org/10.1103/PhysRevA.90.012115}{Phys. Rev. A
  {\bf 90}, 012115}~(2014).

\bibitem{yan2024experimental}
Wen-Zhe Yan, Yunting Li, Zhibo Hou, Huangjun Zhu, Guo-Yong Xiang, Chuan-Feng
  Li, and Guang-Can Guo.
\newblock ``Experimental demonstration of inequivalent mutually unbiased
  bases''.
\newblock \href{https://dx.doi.org/10.1103/PhysRevLett.132.080202}{Phys. Rev.
  Lett. {\bf 132}, 080202}~(2024).

\bibitem{TothG05}
G\'eza T\'oth and Otfried G\"uhne.
\newblock ``Detecting genuine multipartite entanglement with two local
  measurements''.
\newblock \href{https://dx.doi.org/10.1103/PhysRevLett.94.060501}{Phys. Rev.
  Lett. {\bf 94}, 060501}~(2005).

\bibitem{Bavaresco2018}
Jessica Bavaresco, Natalia Herrera~Valencia, Claude Kl{\"o}ckl, Matej
  Pivoluska, Paul Erker, Nicolai Friis, Mehul Malik, and Marcus Huber.
\newblock ``Measurements in two bases are sufficient for certifying
  high-dimensional entanglement''.
\newblock \href{https://dx.doi.org/10.1038/s41567-018-0203-z}{Nat. Phys. {\bf
  14}, 1032--1037}~(2018).

\bibitem{Bae2022}
Joonwoo Bae, Anindita Bera, Dariusz Chruściński, Beatrix~C Hiesmayr, and
  Daniel McNulty.
\newblock ``How many mutually unbiased bases are needed to detect bound
  entangled states?''.
\newblock \href{https://dx.doi.org/10.1088/1751-8121/acaa16}{J. Phys. A: Math.
  Theor. {\bf 55}, 505303}~(2022).

\bibitem{li2025new}
Xiaodi Li.
\newblock ``A new general quantum state verification protocol by the classical
  shadow method''.
\newblock \href{https://dx.doi.org/10.1007/s11128-025-04889-1}{Quantum Inf.
  Process. {\bf 24}, 276}~(2025).

\bibitem{gupta2025}
Meghal Gupta, William He, and Ryan O'Donnell.
\newblock ``Few single-qubit measurements suffice to certify any quantum
  state''~(2025).
\newblock  \href{http://arxiv.org/abs/2506.11355}{arXiv:2506.11355}.

\bibitem{GonzalezAvella2025cyclicmeasurements}
Victor Gonzalez~Avella, Jakub Czartowski, Dardo Goyeneche, and Karol
  {\.{Z}}yczkowski.
\newblock ``Cyclic measurements and simplified quantum state tomography''.
\newblock \href{https://dx.doi.org/10.22331/q-2025-06-04-1763}{{Quantum} {\bf
  9}, 1763}~(2025).

\bibitem{Iosue2024projectivetoric}
Joseph~T. Iosue, T.~C. Mooney, Adam Ehrenberg, and Alexey~V. Gorshkov.
\newblock ``Projective toric designs, quantum state designs, and mutually
  unbiased bases''.
\newblock \href{https://dx.doi.org/10.22331/q-2024-12-03-1546}{{Quantum} {\bf
  8}, 1546}~(2024).

\end{thebibliography}

\onecolumn
\appendix
\numberwithin{equation}{section}
\renewcommand{\theequation}{\thesection\arabic{equation}}

\section*{Appendix}

In this Appendix,  we provide additional information on many variants of the SD protocol and MUB protocol. To evaluate and compare the performance of these verification protocols, we discuss in more detail the verification of GHZ states and provide extensive numerical results on the spectral gaps and sample costs associated with these verification protocols applied to Haar-random pure states, Dicke state, and W states. In addition, we offer some preliminary results on verification protocols beyond MUB. Furthermore, we summarize the distinctions between these verification protocols and the protocol proposed by Huang, Preskill, and Soleimanifar (HPS) \cite{huang2025certifying}. For the convenience of the readers, the sample sizes of Haar-random pure states employed in numerical simulations are summarized in a table at the end of this Appendix.

\section{SD protocol} \label{app:SDprotocol}

In this section we discuss the applications of the SD protocol to the verification of GHZ states and Dicke states. 

\subsection{Verification of GHZ states} \label{app:GHZveriSD}

Here we clarify the performance of the SD protocol applied to the $n$-qudit GHZ state. It turns out that the SD protocol with uniform probability can only achieve spectral gap $2^{1-n}$, which saturates the lower bound in \thref{thm:QSVSDgap} and decreases exponentially with $n$. Nevertheless, the SD protocol with optimized probabilities can achieve spectral gap $1/2$, which is independent of the local dimension and qudit number.

Recall that the $n$-qudit GHZ state in $\caH=\caH_d^{\otimes n}$ can be expressed as follows:
\begin{equation}
	|\rmG_d^n\> := \frac{1}{\sqrt{d}}\sum_{i\in \bbZ_d} |i\>^{\otimes n},  \label{eq:GHZ}
\end{equation}
which can be abbreviated as $|\rmG\>$ when there is no danger of confusion.
Here the computational basis for each party is also a Schmidt basis and the eigenbasis of the generalized Pauli operator $Z = \sum_{i\in \bbZ_d} \omega^i |i\>\<i|$. In addition, the Fourier basis coincides with the eigenbasis of the generalized Pauli operator $X = \sum_{i\in \bbZ_d} |i+1\>\<i|$  and is composed of the following kets:
\begin{equation}
	|\tilde{a}_i\> = \frac{1}{\sqrt{d}}\sum_{j\in \bbZ_d} \omega^{ij} |j\>,\quad i\in \bbZ_d.  \label{eq:FourierMUB}
\end{equation}
By virtue of the Fourier basis, the GHZ state $|\rmG\>$ can also be expressed  as follows: 
\begin{equation}
	|\rmG\> = \frac{1}{\sqrt{d}}\sum_{i\in \bbZ_d}  |\tilde{a}_i\>  \otimes |\tilde{B}_i\>,  \quad
	|\tilde{B}_i\>=\frac{1}{\sqrt{d}}\sum_{j\in \bbZ_d} \omega^{-ij} |j\>^{\otimes (n-1)}.
\end{equation}
Note that $|\tilde{B}_i\>$ for $i\in \bbZ_d$ are GHZ-like states. 

In general, when the SD protocol is applied to the GHZ state $|\rmG\>$, each conditional reduced state that may appear in the verification procedure is either a product state or a GHZ-like state. In addition, the measurement basis of each party (including the last party) is the eigenbasis of either $Z$ or $X$. In this special case, the SD protocol can actually be realized by nonadaptive Pauli $Z$ and $X$ measurements. If all the first $n-1$ parties perform  $X$ measurements, then the last party performs an $X$ measurement as well; otherwise, the last party performs a $Z$ measurement. The $2^{n-1}$ tests featured in the SD protocol can be labeled by strings in $\bbZ_2^{n-1}$ as in the general case. Here we are particularly interested in the two tests labeled by $\mathbf{0}=0^{n-1}$ ($Z$ measurements for all parties) and $\mathbf{1}=1^{n-1}$ ($X$ measurements for all parties); the two corresponding test projectors read
\begin{equation}
    P_{\mathbf{0}} =\sum_{i\in \bbZ_d} (|i\>\<i|)^{\otimes n},\quad P_{\mathbf{1}} =\frac{1}{d}\sum_{i\in \bbZ_d}\left(X^{\otimes n}\right)^i,
\end{equation}
which have ranks $d$ and $d^{n-1}$, respectively. In addition, we have   
\begin{equation}
    \tr(P_{\mathbf{0}} P_{\mathbf{1}})=1,\quad    
    P_\bfm\geq P_{\mathbf{0}} \quad \forall\, \bfm\in \bbZ_2^{n-1}\setminus \{\mathbf{1}\}. 
\end{equation}
 Let $\bP_{\mathbf{0}}=P_{\mathbf{0}}-|\rmG\>\<G|$ and $\bP_{\mathbf{1}}=P_{\mathbf{1}}-|\rmG\>\<G|$; then $\bP_{\mathbf{0}}$ and $\bP_{\mathbf{1}}$ are  projectors that are mutually orthogonal. 
If all tests $P_\bfm$ for $\bfm\in \bbZ_2^{n-1}$ are performed  with uniform probability, then 
\begin{equation}
 \Omega=2^{1-n}\sum_{\bfm \in \bbZ_2^{n-1}} P_\bfm\geq \frac{2^{n-1}-1}{2^{n-1}} P_{\mathbf{0}}+\frac{1}{2^{n-1}}P_{\mathbf{1}},
\end{equation}
which implies that $\nu(\Omega)\leq 2^{1-n}$. In conjunction with \thref{thm:QSVSDgap} we can deduce that $\nu(\Omega)= 2^{1-n}$, which saturates the lower bound in \thref{thm:QSVSDgap}. 

Next, we show that the spectral gap achieved by the SD protocol can be enhanced exponentially if the test probabilities are optimized. Suppose we perform the test $P_\bfm$ with probability $p_\bfm$ with $\sum_{\bfm\in \bbZ_2^{n-1}}p_\bfm=1$. Then
\begin{equation}
 \Omega=\sum_{\bfm \in \bbZ_2^{n-1}} p_\bfm P_\bfm\geq (1-p_\mathbf{1}) P_{\mathbf{0}}+p_\mathbf{1}P_{\mathbf{1}},
\end{equation}
which implies that
\begin{equation}
\nu(\Omega)\leq \min\{1-p_\mathbf{1}, p_\mathbf{1}\}\leq \frac{1}{2}.
\end{equation}
Here the upper bound is saturated when $p_\mathbf{1}=p_\mathbf{0}=1/2$ and $p_\bfm=0$ for all $\bfm\in \bbZ_2^{n-1}\setminus\{\mathbf{0},\mathbf{1}\}$. 

\begin{figure}
	\centering
	\includegraphics[scale=0.37]{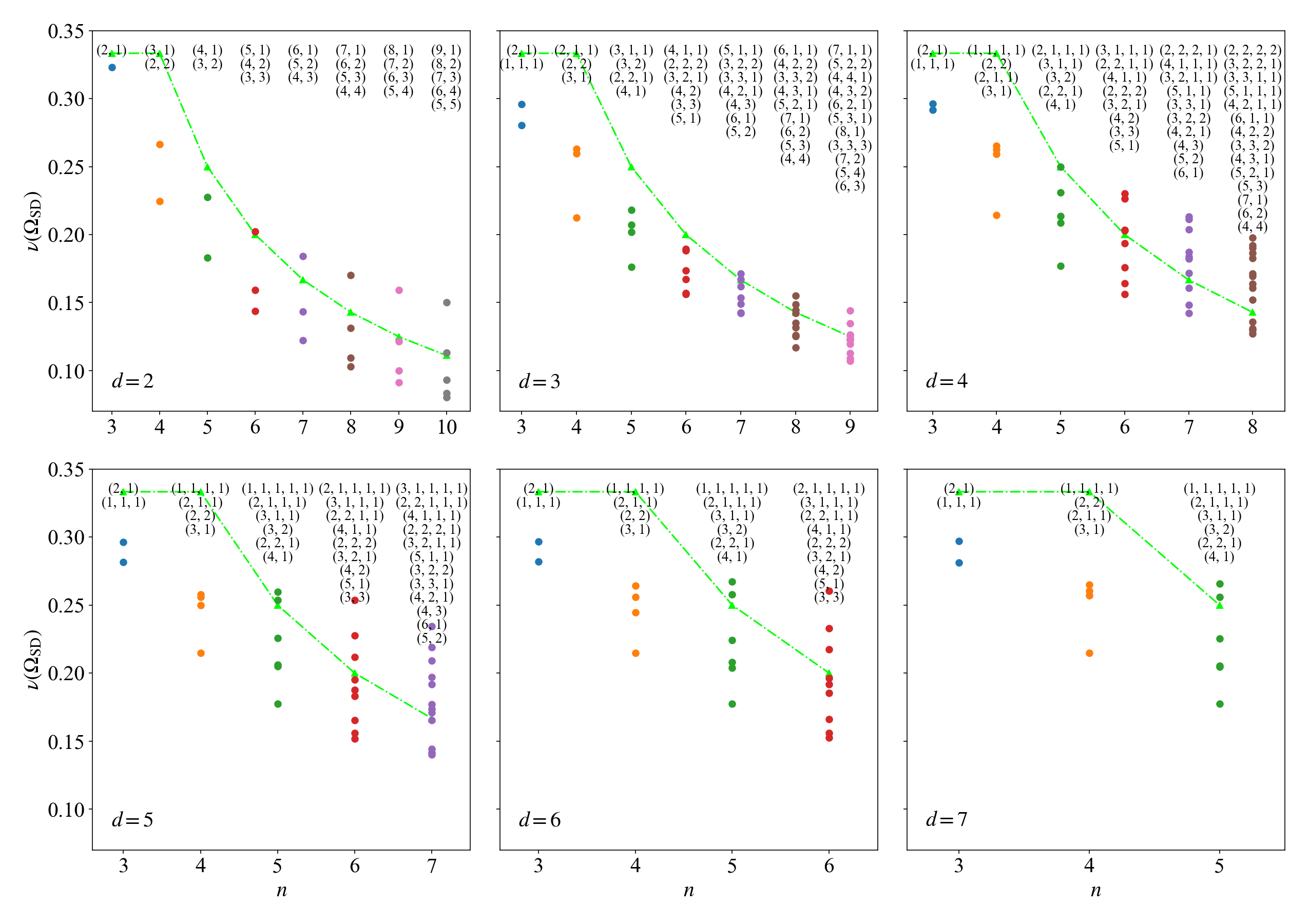}
    \vspace{-1.5ex}
	\caption{The spectral gaps $\nu(\Omega_{\SD})$ achieved by the SD protocol (under the uniform probability distribution) for various Dicke states. For each plot,  the local dimension $d$ is fixed as labeled in the plot. Each data point corresponds to a particular Dicke state $|\rmD_d^n(\bfn)\>$ with the label $\bfn$  listed at the top according to the value of the spectral gap. The triangles on the green dash-dot line represent the spectral gaps achieved by  a specialized verification protocol  as reproduced in \eref{eq:Omegabfn} \cite{LiuYSZ19,Li2021Dicke}.}
	\label{fig:DickeGapn}
\vspace{3ex}
    \includegraphics[scale=0.38]{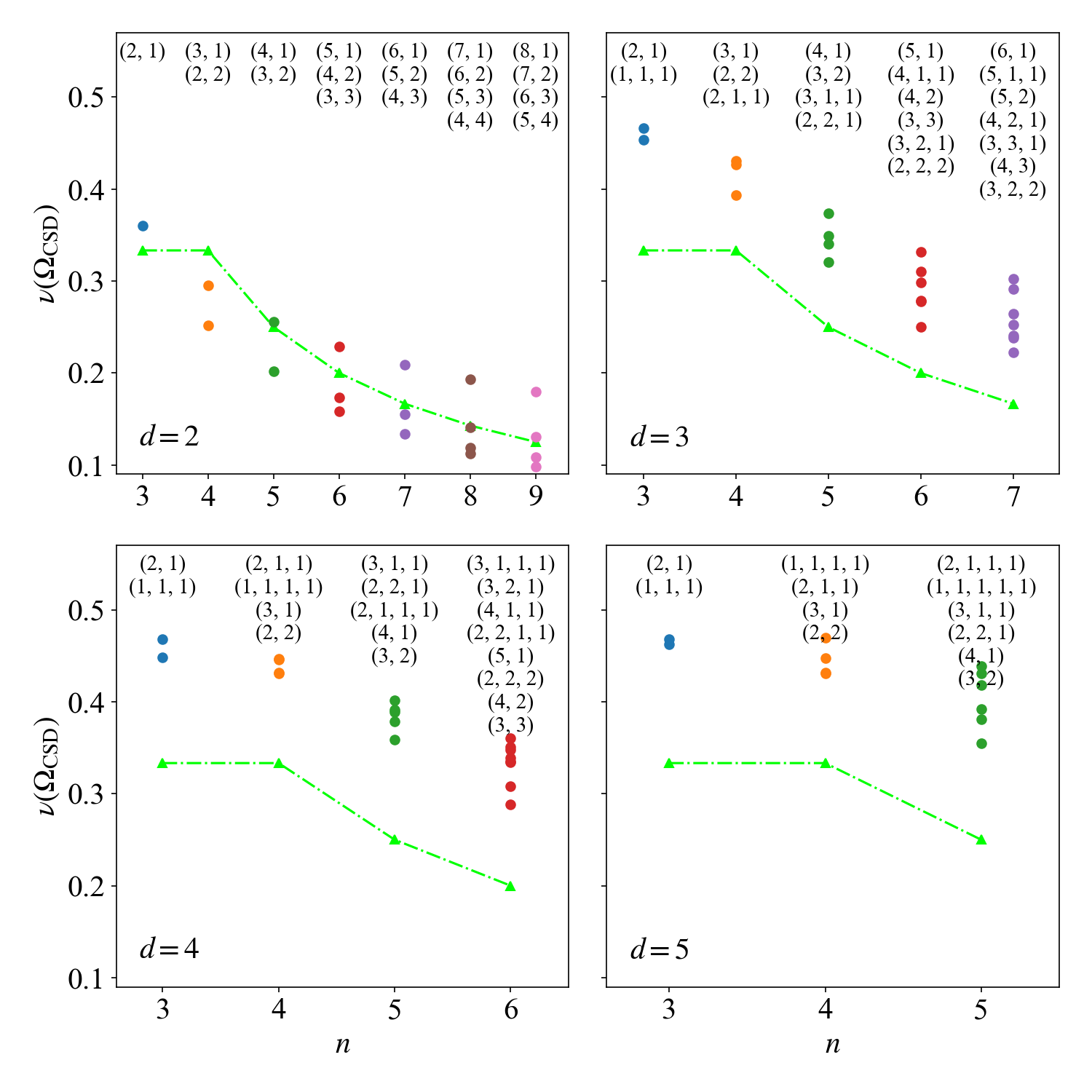}
    \vspace{-1.5ex}
	\caption{
		The spectral gaps $\nu(\Omega_\CSD)$ achieved by the CSD protocol (under the uniform probability distribution) for various  Dicke states. Each data point corresponds to a particular Dicke state $|\rmD_d^n(\bfn)\>$ with the label $\bfn$  listed at the top according to the value of the spectral gap. The triangles on the green dash-dot line represent the spectral gaps determined by  \eref{eq:Omegabfn} as shown in \fref{fig:DickeGapn}.}
	\label{fig:Dicke_CSD}
\end{figure}

\subsection{Verification of Dicke states} \label{app:Dicke}

Dicke states are an important family of multipartite quantum states that are useful in many tasks in quantum information processing, including quantum communication and
quantum metrology. Efficient verification protocols tailored to Dicke states have been constructed recently \cite{LiuYSZ19,Li2021Dicke}. Here we shall illustrate the performance of our universal SD protocol applied to Dicke states in comparison with a specialized protocol proposed in \rcite{Li2021Dicke}, which generalizes a protocol proposed in \rcite{LiuYSZ19}. 

Up to  local unitary transformations, each $n$-qudit Dicke state can be labeled by a nonincreasing sequence of $\ell+1$ ordered positive integers that sum up to $n$, where $1\leq \ell \le \min\{d,n\}-1$. Denote by $\caS_d^n$ the set of all such sequences for  given $n$ and $d$. Consider a general sequence in $\caS_d^n$:
\begin{equation}
    \bfn = (n_0, n_1, \dots, n_\ell),\quad n_0 \ge n_1 \ge \dots \ge n_\ell\geq 1,\quad
	\sum_{i=0}^{\ell} n_i=n;   \label{eq:DickenCondition}
\end{equation}
let $\bbZ_d$ be the ring of integers modulo $d$ and $\scrB(\bfn)$ the set of all strings in $\bbZ_d^n$
in which $n_i$ symbols are equal to $i$ for $i=0,1,\dots,\ell$. Then
the $n$-qudit Dicke state associated with the sequence $\bfn$ is defined as follows \cite{Li2021Dicke}:
\begin{equation}
    |\rmD_d^n(\bfn)\> := \frac{1}{\sqrt{|\scrB(\bfn)|}} \sum_{\bfj \in  \scrB(\bfn)} |\bfj\>,  \label{eq:DickeState}
\end{equation}
where
\begin{equation}
    |\scrB(\bfn)|=\frac{n!}{\prod_{i=0}^{\ell}n_i!}.
\end{equation}
When $n=4$ and $d=3$ for example, $\ell$ can take on two different values, namely, $\ell=1,2$, and the set $\caS_d^n$ is composed of the following three sequences:
\begin{equation}
  (3,1),\quad (2,2),\quad (2,1,1). 
\end{equation}
If  $\bfn=(2,1,1)$, then $\scrB(\bfn)$ is composed of 12 strings in $\bbZ_3^4$:
\begin{equation}
\scrB(\bfn)=\{0012, 0021, 0102, 0120, 0201, 0210, 
1002, 1020, 1200, 2001, 2010, 2100\}, 
\end{equation}
so $|\rmD_d^n(\bfn)\>$ is a superposition of 12 computational basis states.

Next, we illustrate the performance of the SD protocol in verifying various Dicke states, assuming that all tests are chosen with uniform probability. The resulting spectral gaps $\nu(\Omega_{\SD})$ are shown in \fref{fig:DickeGapn}.  As a benchmark, the figure also shows  the spectral gaps achieved by a specialized verification protocol, which are characterized by the formula \cite{LiuYSZ19,Li2021Dicke}:
\begin{equation}
    \nu(\Omega_\bfn) = 
    \begin{cases}
    1/2 & \bfn=(1,1,1), \\
    1/3 & \bfn=(2, 1), \\
    1/(n-1) & n \ge 4.
    \end{cases}  \label{eq:Omegabfn}
\end{equation}
Overall, the SD protocol is comparable to  this specialized  protocol  with respect to the spectral gap (and sample complexity as well); in addition,
for many Dicke states, the SD protocol is much better. These results highlight the versatility and power of the SD protocol in QSV.

\section{CSD protocol} \label{app:CSDprotocol}

The CSD protocol is a probabilistic mixture of $n$ SD protocols associated with $n$ different orders of SD along a cyclic sequence, as illustrated in \fref{fig:CSDframework} in the main text. To demonstrate the performance of the CSD protocol (under the uniform probability distribution), here we take Dicke states as an example.
The spectral gaps $\nu(\Omega_\CSD)$  for various $n$-qudit Dicke states are shown in \fref{fig:Dicke_CSD}. Compared with the SD protocol, the CSD protocol can increase the spectral gaps by about 9.2\% to 83.5\% depending on the specific Dicke state in addition to the values of $d$ and $n$. See \aref{app:compareProtocols} for additional discussion in terms of the sample costs.

\begin{figure}[tbp]
	\centering
	\includegraphics[scale=0.41]{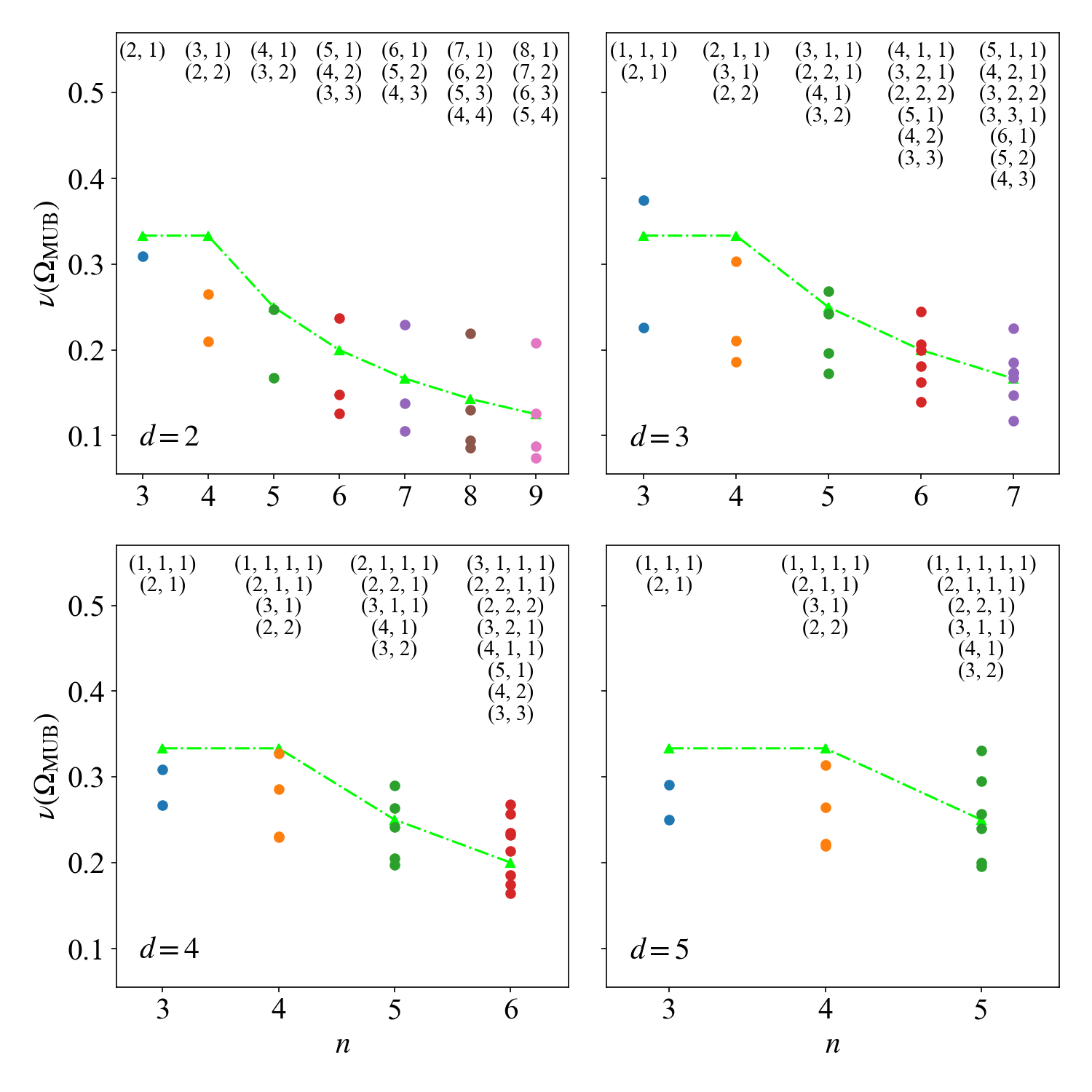}
      \vspace{-1.5ex}
	\caption{
		The spectral gaps $\nu(\Omega_\MUB)$ achieved by the MUB protocol (under the uniform probability distribution) for various  Dicke states. Each data point corresponds to a particular Dicke state $|\rmD_d^n(\bfn)\>$ with the label $\bfn$  listed at the top according to the value of the spectral gap. The triangles on the green dash-dot line represent the spectral gaps determined by  \eref{eq:Omegabfn} as shown in \fref{fig:DickeGapn}. }
	\label{fig:DickeMUB2B}
\vspace{3ex}
\includegraphics[scale=0.41]{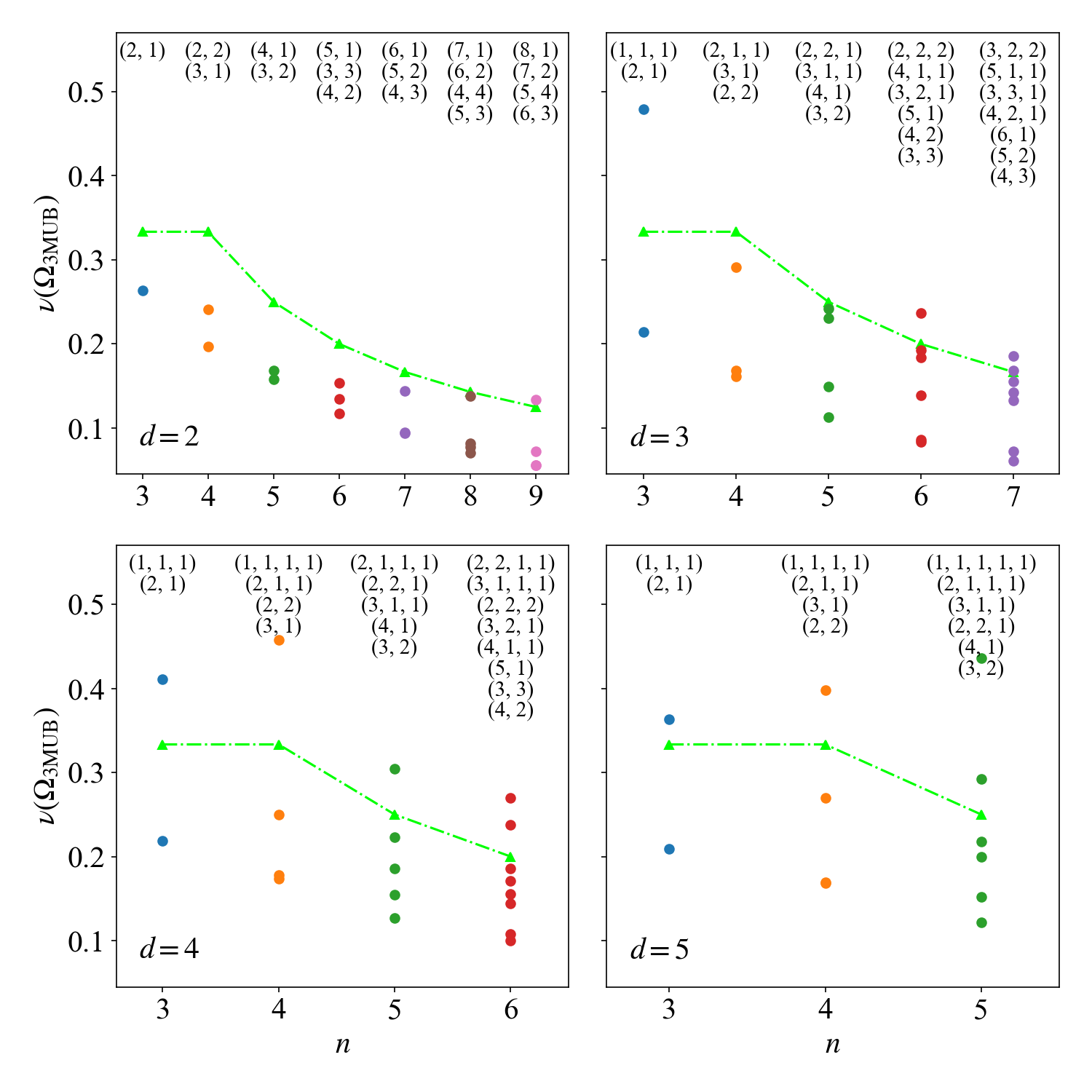}
  \vspace{-1.5ex}
	\caption{
		The spectral gaps $\nu(\Omega_\mathrm{3MUB})$ achieved by the 3MUB protocol (under the uniform probability distribution) for various  Dicke states. Each data point corresponds to a particular Dicke state $|\rmD_d^n(\bfn)\>$ with the label $\bfn$  listed at the top according to the value of the spectral gap. The triangles on the green dash-dot line represent the spectral gaps determined by  \eref{eq:Omegabfn} as shown in \fref{fig:DickeGapn}. }
	\label{fig:DickeMUB3B}
\end{figure}

\begin{figure}[tbp]
	\centering
	\includegraphics[scale=0.41]{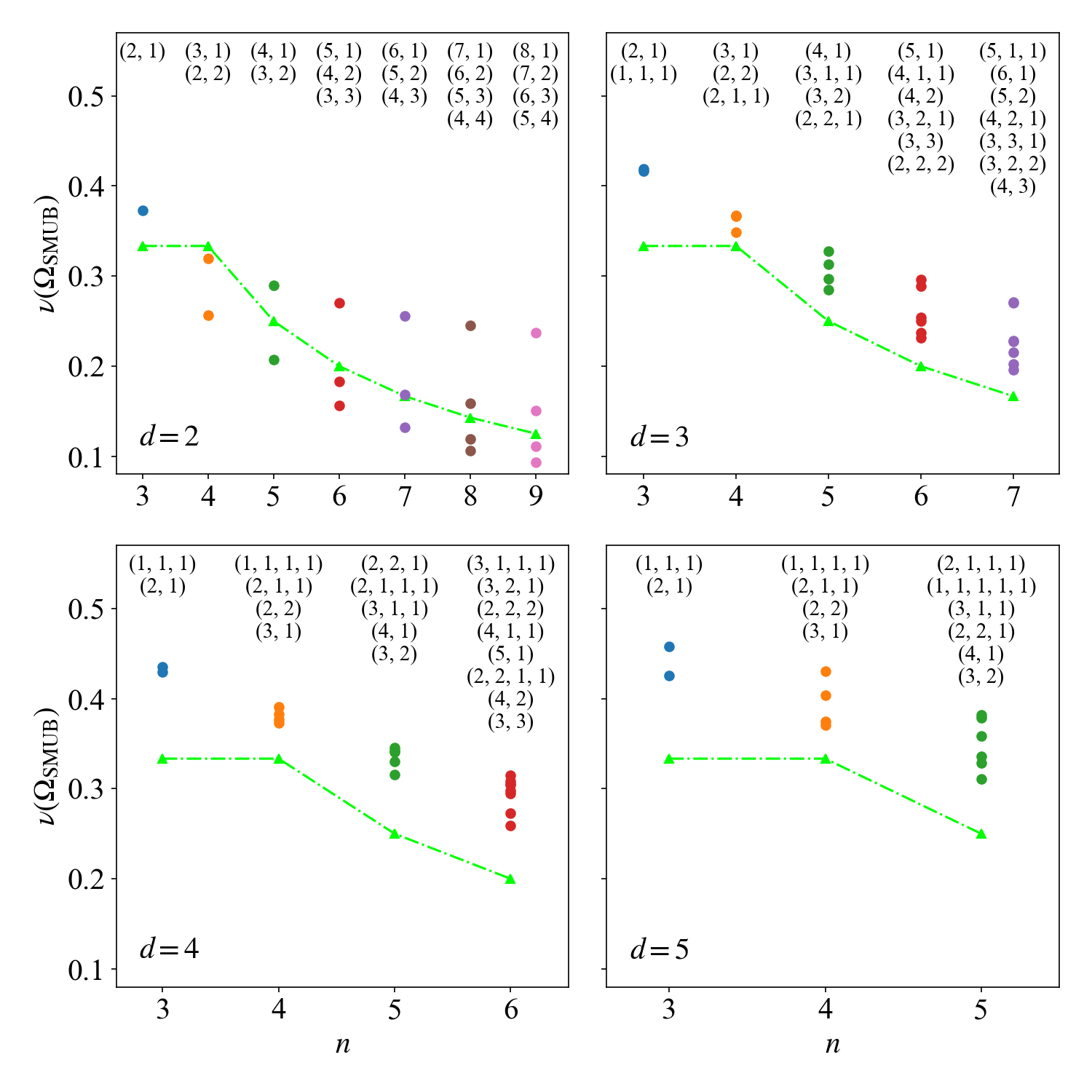}
      \vspace{-1.5ex}
	\caption{
		The spectral gaps $\nu(\Omega_\SMUB)$ achieved by the SMUB protocol (under the uniform probability distribution) for various  Dicke states. Each data point corresponds to a particular Dicke state $|\rmD_d^n(\bfn)\>$ with the label $\bfn$  listed at the top according to the value of the spectral gap. The triangles on the green dash-dot line represent the spectral gaps determined by  \eref{eq:Omegabfn} as shown in \fref{fig:DickeGapn}.  }
	\label{fig:DickeSMUB2B}
\vspace{3ex}
    \includegraphics[scale=0.41]{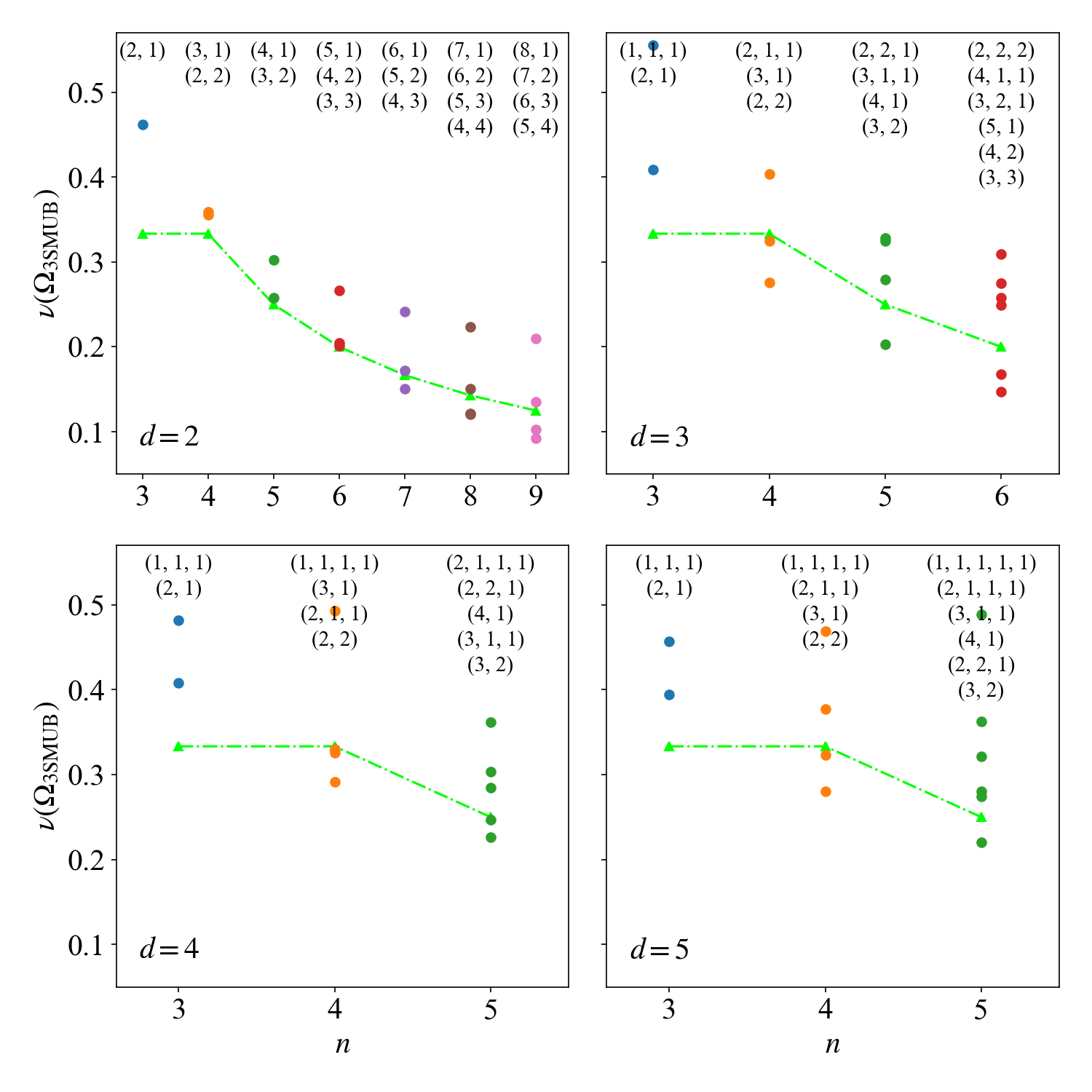}
    \vspace{-1.5ex}
\caption{
The spectral gaps $\nu(\Omega_\mathrm{3SMUB})$  achieved by the 3SMUB protocol (under the uniform probability distribution) for various  Dicke states. Each data point corresponds to a particular Dicke state $|\rmD_d^n(\bfn)\>$ with the label $\bfn$  listed at the top according to the value of the spectral gap. The triangles on the green dash-dot line represent the spectral gaps determined by  \eref{eq:Omegabfn} as shown in \fref{fig:DickeGapn}. }
\label{fig:Dicke3SMUB}
\end{figure}

\section{MUB and CMUB protocols} \label{app:MUBCMUBprotocols}

In this section we illustrate the performance of the MUB and CMUB protocols and their variants in verifying GHZ states and Dicke states. 

\subsection{Verification of GHZ states}

Here we clarify the performance of the MUB protocol applied to the $n$-qudit GHZ state
$|\rmG_d^n\>$ in \eref{eq:GHZ}. For concreteness, we focus on the MUB protocol based on generalized Pauli $Z$ and $X$ measurements. Note that the eigenbasis of $Z$ coincides with the computational basis, while the eigenbasis of $X$ coincides with the Fourier basis presented in \eref{eq:FourierMUB}. 
For the GHZ state $|\rmG_d^n\>$, therefore, the MUB protocol is equivalent to the SD protocol (see \aref{app:GHZveriSD}). Accordingly, the MUB protocol with uniform test probabilities can only achieve spectral gap $2^{1-n}$, while the MUB protocol with optimized probabilities can achieve spectral gap $1/2$, which is independent of the local dimension and qudit number. Note that the optimal MUB protocol consists of only two distinct tests and coincides with the CMUB protocol with uniform test probabilities.

If the local dimension $d$ is a prime,  then the respective eigenbases of $X, XZ, XZ^2, \dots, XZ^{d-1}, Z$ form a complete set of MUB as mentioned in the main text, from which we can construct a cMUB protocol. For the $n$-qudit GHZ state, some variant of the cMUB protocol was studied earlier. According to \rcite{Li2020GHZ}, the cMUB protocol with optimized test probabilities can achieve spectral gap $d/(d+1)$, which is the maximum spectral gap achievable by local operations and classical communication and increases monotonically with the local dimension $d$.

\subsection{Verification of  Dicke states}
Here we illustrate the performance of the MUB and 3MUB protocols (under  uniform probability distributions) in verifying Dicke states. 
The spectral gaps $\nu(\Omega_\MUB)$ and $\nu(\Omega_\mathrm{3MUB})$ for various  Dicke states are shown in \fsref{fig:DickeMUB2B} and \ref{fig:DickeMUB3B}, respectively. The performance of both protocols is comparable to the SD protocol and a specialized verification protocol proposed in \rscite{LiuYSZ19,Li2021Dicke}.

\section{SMUB and SCMUB protocols} \label{app:SMUBSCMUBprotocols}
In this section we first discuss the construction of test projectors and verification operators of the SMUB and SCMUB protocols. Then we illustrate the performance of the SMUB and SCMUB protocols and their variants in verifying various Dicke states and the impact of probability optimization on the spectral gaps.

\begin{figure}[bt]
	\centering
	\includegraphics[scale=0.4]{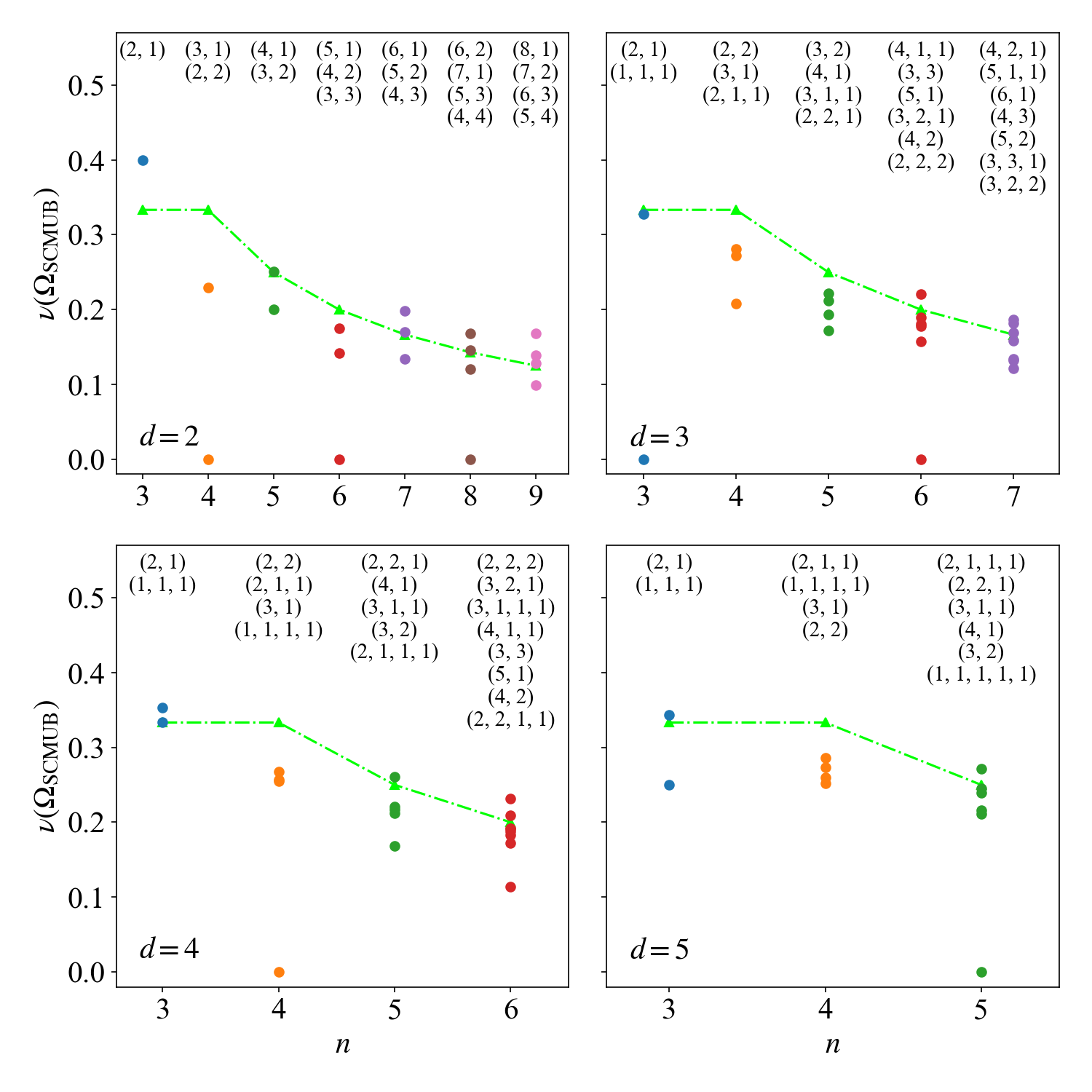}
    \vspace{-1.5ex}
	\caption{
		The spectral gaps $\nu(\Omega_\mathrm{SCMUB})$  achieved by the SCMUB protocol (under the uniform probability distribution) for various  Dicke states. Each data point corresponds to a particular Dicke state $|\rmD_d^n(\bfn)\>$ with the label $\bfn$  listed at the top according to the value of the spectral gap. The triangles on the green dash-dot line represent the spectral gaps determined by  \eref{eq:Omegabfn} as shown in \fref{fig:DickeGapn}. }
	\label{fig:DickeSCMUB}
\end{figure}

\subsection{Test projectors and verification operators featured in the SMUB and SCMUB protocols}

Recall that the SMUB protocol is constructed by randomizing the choice of the last party in the MUB protocol, as illustrated  in \fref{fig:SMUBframework} in the main text. It is composed of $2^{n-1}n$ tests, and each test is labeled by the choice of the last party $k$ and a string $\bfm$ in $\bbZ_2^{n-1}$, which specifies the measurement setting of the other $n-1$ parties. In analogy to  \eref{eq:QmGPauli}, the corresponding test projector can be expressed as follows:
\begin{equation}
    Q_\bfm^k 
    = \sum_{\bfo \in \bbZ_d^{n-1}} \Pi_{o_1, m_1}\otimes \cdots  
   \otimes \Pi_{o_{k-1}, m_{k-1}} 
     \otimes|\psi_{k,\bfo, \bfm}\>\<\psi_{k,\bfo, \bfm}| 
     \otimes \Pi_{o_{k+1}, m_{k+1}} \otimes \cdots   \otimes \Pi_{o_{n}, m_{n}},  
\end{equation}
where $\Pi_{o_l, m_l}$ for each $l\in \{1,2,\ldots, n\}\setminus \{k\}$ denotes the projector onto the $o_l$-th basis state associated with the computational basis (when $m_l=0$) or MUB (when $m_l=1$),
and $|\psi_{k,\bfo, \bfm}\>$ is the conditional reduced state of the target state $|\Psi\>$ for party $k$ given the measurement setting $\bfm$ and outcome $\bfo$.

Suppose party $k$ is chosen with probability $p_k$ and, for a given $k$, the test  $Q_\bfm^k$ is chosen with probability $q_\bfm^k$, assuming that $\sum_{k=1}^n p_k=1$ and $\sum_{\bfm\in \bbZ_2^{n-1}} q_\bfm^k=1$ for each $k$. Then the verification operator associated with party $k$ reads
\begin{equation}
\Omega_{\MUB,k} = \sum_{\bfm \in \bbZ_2^{n-1}} q_\bfm^k Q_\bfm^k,  \label{eq:OmegaMUBk}
\end{equation}
and the overall  verification operator reads
\begin{equation}
    \Omega_{\SMUB} = \sum_{k=1}^{n}  p_k \Omega_{\MUB,k}= \sum_{k=1}^{n} \sum_{\bfm \in \bbZ_2^{n-1}}  p_k q_\bfm^k Q_\bfm^k 
    = \sum_{k=1}^{n} \sum_{\bfm \in \bbZ_2^{n-1}}  \tilde{q}_\bfm^k Q_\bfm^k,
\end{equation}
where $\tilde{q}_\bfm^k=p_k q_\bfm^k$. If all tests are chosen with the same probability, then 
\begin{equation} 
\Omega_{\MUB,k} = 2^{1-n}\sum_{\bfm \in \bbZ_2^{n-1}} Q_\bfm^k, \quad
\Omega_{\SMUB} = \frac{1}{n}\sum_{k=1}^{n}  \Omega_{\MUB,k}= \frac{2}{2^n n}\sum_{k=1}^{n} \sum_{\bfm \in \bbZ_2^{n-1}}  Q_\bfm^k.  \label{eq:OmegaSMUB}
\end{equation}

The test projectors and verification operators of the 3SMUB, SCMUB, and 3SCMUB protocols (and other variants) can be constructed in a similar way. Note that the 3SMUB protocol is composed of $3^{n-1}n$ tests, and each test $Q_\bfm^k$ is labeled by the choice of the last party $k$ and a string $\bfm$ in $\bbZ_3^{n-1}$, which specifies the measurement setting of the other $n-1$ parties. The test set of the SCMUB (3SCMUB) protocol is contained in the counterpart of the SMUB (3SMUB) protocol.

\subsection{Verification of  Dicke states}

The spectral gaps achieved by the SMUB protocol (under the uniform probability distribution) for various Dicke states are shown in 
\fref{fig:DickeSMUB2B} in comparison with the spectral gaps achieved by a specialized verification protocol as determined by \eref{eq:Omegabfn} \cite{LiuYSZ19,Li2021Dicke}.  The counterparts for the 3SMUB and SCMUB protocols 
are shown in \fsref{fig:Dicke3SMUB} and \ref{fig:DickeSCMUB}, respectively. These figures indicate that the SMUB and 3SMUB protocols  can verify all Dicke states, while the SCMUB protocol can verify most Dicke states.

\subsection{Test probability optimization of  the 3SMUB protocol for $n$-qubit Dicke states}

\begin{figure}[!t]
	\centering
	\includegraphics[scale=0.45]{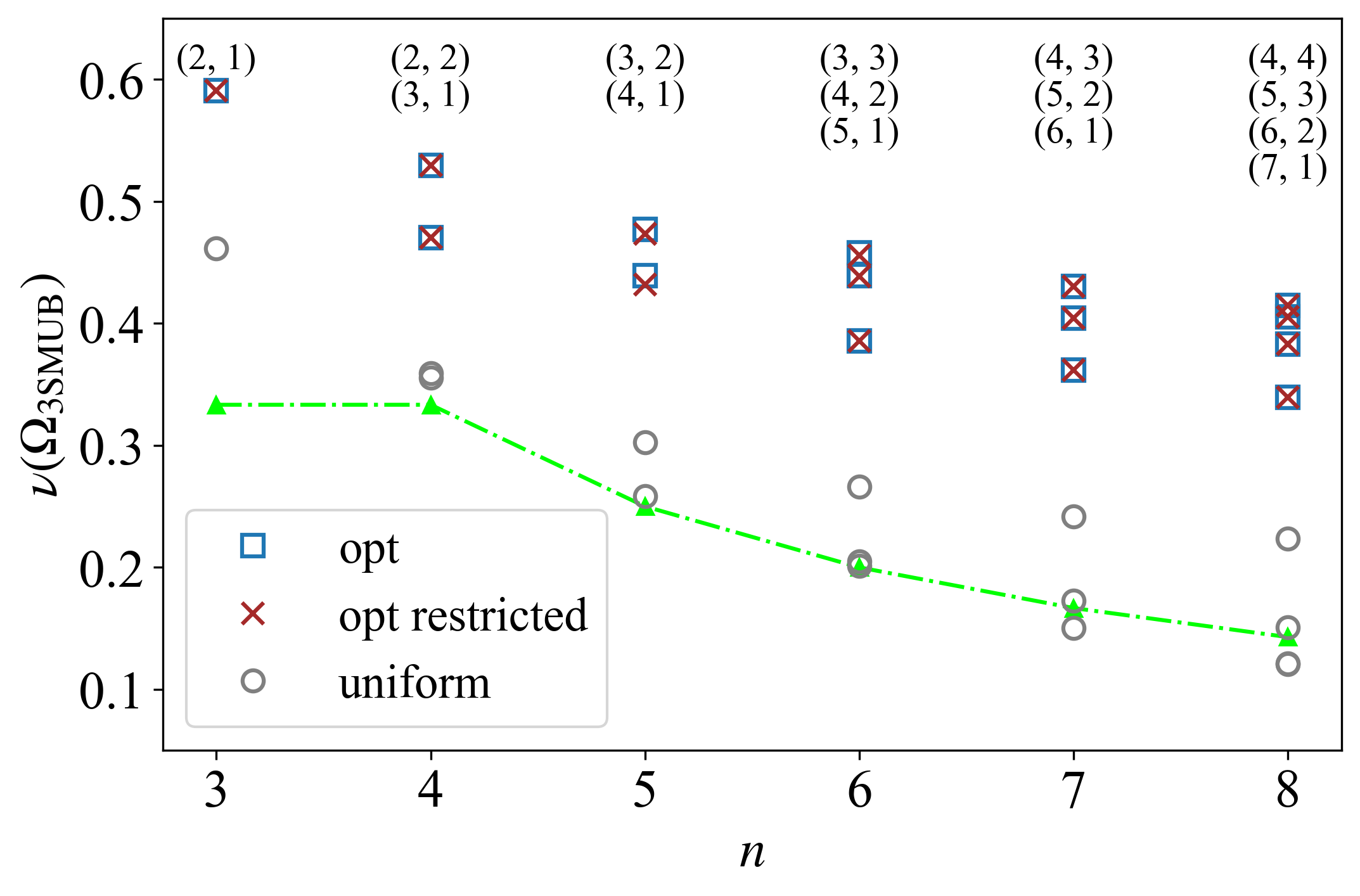}
    \vspace{-1ex}
	\caption{The spectral gaps $\nu(\Omega_{\mathrm{3SMUB}})$ achieved by  three variants of the 3SMUB protocol for $n$-qubit Dicke states. Two variants are based on the uniform and optimal (opt) probability distributions, respectively, while the third one is tied to the restricted 3SMUB protocol characterized by \eref{eq:OmegaDickeOptSymp} with the optimized probability distribution (opt restricted). Each data point corresponds to a particular Dicke state $|\rmD_d^n(\bfn)\>$ with the label $\bfn$  listed at the top according to the value of the spectral gap. The triangles on the green dash-dot line represent the spectral gaps determined by  \eref{eq:Omegabfn} as shown in \fref{fig:DickeGapn}. }
	\label{fig:Dicke3SMUBd2}
\end{figure}

In this section we study the impact of probability optimization on the 3SMUB protocol for verifying $n$-qubit Dicke states. For a given Dicke state, the maximum spectral gap and the optimal probabilities for performing individual tests can be determined by semidefinite programming (SDP) as shown in Eq.~(3). The maximum spectral gaps for various Dicke states up to eight qubits are shown in \fref{fig:Dicke3SMUBd2} in comparison with the counterparts based on the uniform probability distributions. According to \fref{fig:Dicke3SMUBd2}, probability optimization can usually more than double the spectral gaps although it is not so useful for Haar-random pure states. The optimal probabilities for two Dicke states labeled by the two sequences $\bfn=(2, 2)$ and $\bfn=(4, 1)$ are shown in \tref{tab:DickeSDP}.

\begin{table}[t]
\renewcommand{\arraystretch}{1.23}
\setlength{\tabcolsep}{7pt}
\centering
\caption{Optimized test probabilities $q_\bfm^k$ 
for applying the 3SMUB protocol to verifying the two Dicke states labeled by $\bfn=(2, 2)$ and $\bfn=(4, 1)$, respectively. Note that $p_k=1/n$ and $\tilde{q}_\bfm^k=q_\bfm^k/n$ for $k=1,2,\ldots, n$. Here $\caP[\bfm]$ denotes the set of all distinct permutations of the string $\bfm$; $n_z$ is the number of entries in $\bfm$ that are equal to 0, which corresponds to  the number of parties that perform Pauli $Z$ measurements. Strings that do not appear mean the corresponding probabilities are zero. } \label{tab:DickeSDP}
\begin{tabular} {c|ccc cc }  \hline \hline 
$n$ & $\bfn$ & $n_z$ & $\bfm$  & string number & $q_\bfm^k$ \\
\hline
\multirow{4}{*}{4} & \multirow{4}{*}{$(2, 2)$} & 0 & $(1,1,1), (2,2,2) $ & 2 & 0.0882 \\
 & & 1 & $\caP[(0,1,1)], \caP[(0,2,2)]$ & 6 & 0.0817 \\
 & & 2 & $\caP[(0,0,1)], \caP[(0,0,2)]$ & 6 & 0.0458 \\
 & & 3 & $(0,0,0)$ & 1 & 0.0588 \\
\hline
\multirow{3}{*}{5} & \multirow{3}{*}{$(4, 1)$} & 0 & $(1,1,1,1), (2,2,2,2)$ & 2 & 0.2221 \\
 & & 4 & $(0,0,0,0)$ & 1 & 0.2422 \\
 & & 0 & $\caP[(1,1,2,2)]$ & 6 & 0.0523 \\
\hline \hline
\end{tabular}
\end{table}

When $n$ is large, it is quite resource-intensive to determine the maximum spectral gap by SDP. To alleviate this task, we can group certain test projectors before running SDP. Denote by $\caZ_{n-1}(n_z)$ the set of strings in $\bbZ_3^{n-1}$ in which $n_z$ entries are equal to 0 (corresponding to Pauli $Z$ measurements), and the remaining $n-n_z-1$ entries are all equal to 1 or all equal to 2 (corresponding to Pauli $X$ or $Y$ measurements). The cardinality of this set is $|\caZ_{n-1}(n_z)|=2 \binom{n-1}{n_z}$.
For each $n_z=0,1,\dots,n-1$, 
perform all tests $Q_\bfm^k$ for $k=1,2,\ldots, n$ and $\bfm \in \caZ_{n-1}(n_z)$ with the same probability.  The resulting effective test operator $T_{n_z}$ reads
\begin{equation}
    T_{n_z} = \frac{1}{n|\caZ_{n-1}(n_z)|} \sum_{k=1}^n\sum_{\bfm \in \caZ_{n-1}(n_z)} Q_\bfm^k.
\end{equation}
If $T_{n_z}$ is performed with probability $p_{n_z}$ with  $\sum_{n_z=0}^{n-1} p_{n_z} = 1$, then the resulting verification operator $\Omega$ reads
\begin{equation}
    \Omega = \sum_{n_z=0}^{n-1} p_{n_z} T_{n_z}.  \label{eq:OmegaDickeOptSymp}
\end{equation}
Now, the probabilities $p_{n_z}$ can be  optimized using SDP. 
The resulting spectral gaps $\nu(\Omega_{3\SMUB})$ for various qubit Dicke states are shown  as crosses in \fref{fig:Dicke3SMUBd2}. Surprisingly, the  spectral gaps  achieved by this simplified optimization procedure are quite close to the counterparts based on brute-force optimization. This observation is quite helpful for constructing efficient verification protocols for Dicke states.

\begin{figure}
	\centering
	\includegraphics[scale=0.46]{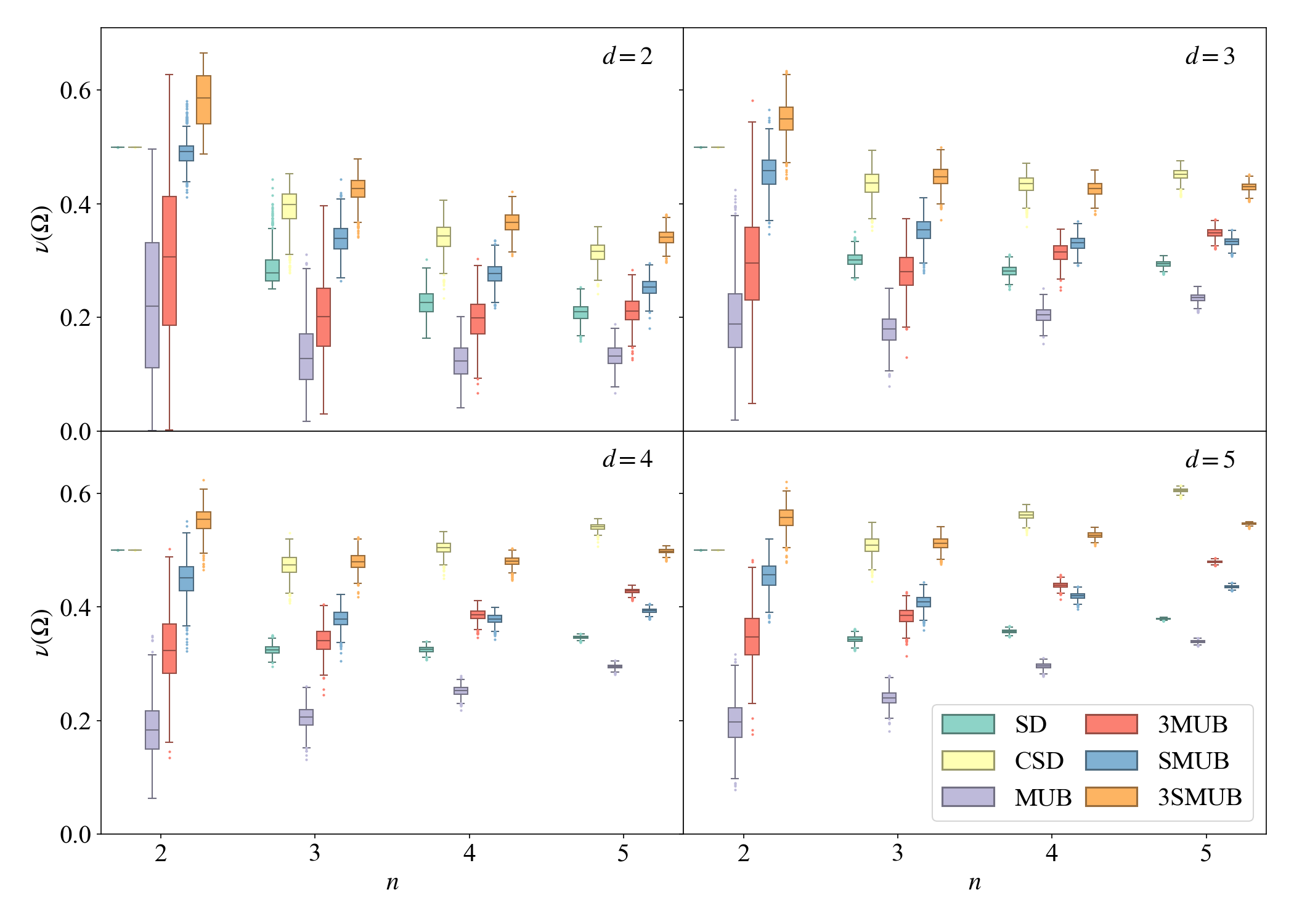}
      \vspace{-2ex}
	\caption{Box plots of the spectral gaps achieved by six variants of the SD and MUB protocols for $n$-qudit Haar-random pure states. For each local dimension $d$ and qudit number $n$, 1000 Haar-random pure states are generated (see \tref{tab:SampleNumHaar}).    
		Each box spans the interquartile range (IQR) of the middle 50\% of the data, with the central line representing the median. The whiskers extend to the minimum and maximum values within $1.5\times$IQR from the lower and upper quartiles, respectively; the points beyond the whiskers are considered outliers.}	
	\label{fig:hboxCompare}
    
\vspace{3.5ex}
\includegraphics[scale=0.56]{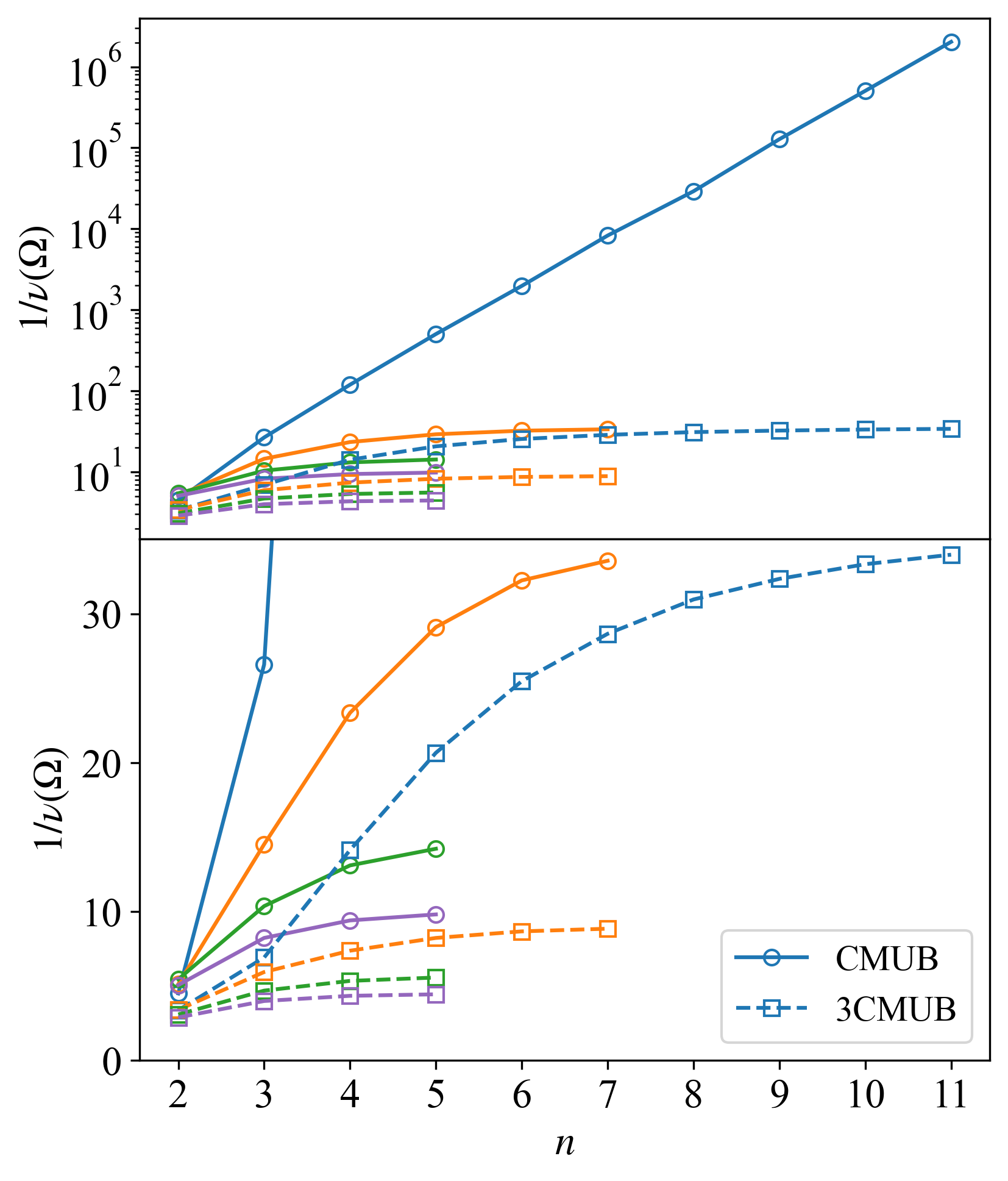}
\vspace{-1ex}
\caption{The inverses of average spectral gaps $1/\nu(\Omega_{\mathrm{CMUB}})$ achieved by the CMUB protocol (solid lines with circles) and 3CMUB protocol (dashed lines with squares) in logarithmic scale (top) and linear scale (bottom). The color encoding of the local dimension $d$ follows \fref{fig:HaarMUB} in the main text (blue, orange, green, and purple mean $d=2,3,4,5$, respectively).}
\label{fig:GapInvCMUB3CMUBlog2}
\end{figure}

\section{Comparison of various verification protocols} \label{app:compareProtocols}
In this section we compare several variants of the SD and MUB protocols with respect to the spectral gaps and sample costs. Haar-random pure states, Dicke states, and W states are considered separately.

\subsection{Comparison for Haar-random pure states}
In the case of Haar-random pure states, we first compare the spectral gaps achieved by eight variants of the SD and MUB protocols, namely, SD, CSD, MUB, 3MUB, SMUB,  3SMUB, SCMUB, and 3SCMUB protocols. The distributions of the spectral gaps achieved by the first six protocols for $n,d=2,3,4,5$ are illustrated as  box plots in \fref{fig:hboxCompare}. For all protocols, the spectral gaps tend to increase with the local dimension $d$ as mentioned in the main text.

\begin{table}[tbp]
\renewcommand{\arraystretch}{1.32}
\setlength{\tabcolsep}{7pt}
\centering
\caption{The reduction rates in sample costs of the CSD, MUB, 3MUB, SMUB, 3SMUB, SCMUB, and 3SCMUB protocols compared with the SD protocol in the verification of $n$-qudit Haar-random pure states. For each local dimension $d$ and qudit number $n$, 1000 Haar-random pure states are generated as in \fref{fig:hboxCompare} (see \tref{tab:SampleNumHaar}).}  \label{tab:compareHaardn}
\begin{tabular}{c|c| cccc ccc} 
\hline \hline 
$d$ & $n$ & CSD \% & MUB \% & 3MUB \% & SMUB \% & 3SMUB \% & SCMUB \% & 3SCMUB \%\\
\hline
\multirow{4}{*}{2} 
 & 2 & 0 & $-695.9$ & $-330.3$ & $-2.2$ &13.4 & $-1.9$ & 14.1 \\
 & 3 & 27.1 & $-172.2$ & $-66.6$ & 15.8 &33.0 & 7.1 & 26.8\\
 & 4 & 34.1 & $-96.0$ & $-19.2$ &  18.7 & 39.0 &2.1 & 26.5 \\
 & 5 & 33.7 & $-60.8$ & $0.2$ & 17.5 & 39.0 & $-5.8$ & 21.3\\
\hline
\multirow{4}{*}{3} 
 & 2 & 0 & $-204.2$ & $-93.0$ & $-10.3$ & 8.7 & $-9.8$ & 9.0\\
 & 3 & 30.6 & $-73.9$ & $-9.8$ & 14.5 & 32.6 & 5.9 & 26.1 \\
 & 4 & 35.0 & $-38.6$ & 10.2 & 14.7 & 33.9 & $-0.9$ & 22.5\\
 & 5 & 34.8 & $-25.5$ & 15.6 & 11.6 & 31.5 & $-9.3$ & 16.2\\
\hline
\multirow{4}{*}{4} 
 & 2 & 0 & $-197.2$ & $-60.7$ & $-12.1$ & 9.3 & $-11.5$ & 9.5\\
 & 3 & 31.4 & $-59.6$ & 4.2 & 14.3 & 32.5 & 7.0 & 26.5\\
 & 4 & 35.5 & $-29.0$ & 15.7 & 14.2 & 32.4 & 1.4 & 22.7 \\
 & 5 & 35.9 & $-17.5$ & 19.0 & 11.8 & 30.4 & $-4.4$ & 18.0\\
\hline
\multirow{4}{*}{5} 
 & 2 & 0 & $-164.0$ & $-47.1$ & $-10.3$ & 10.1 & $-10.0$ & 10.2 \\
 & 3 & 32.4 & $-43.6$ & 10.5 & 15.9 & 33.0  & 9.4 & 27.8 \\
 & 4 & 36.4 & $-20.6$ & 18.5 & 14.9 & 32.2 & 4.0 & 23.8\\
 & 5 & 37.3 & $-11.8$ & 21.0 & 13.0 & 30.7  & 0.5 & 20.7\\
\hline \hline
\end{tabular}

\vspace{5ex}
	\caption{The reduction rates in sample costs of the CSD, MUB, 3MUB, SMUB, and 3SMUB protocols compared with the SD protocol in the verification of $n$-qudit Dicke states.} \label{tab:compareDickedn}
	\begin{tabular}{c|c| ccc cc}   
		\hline \hline 
		$d$ & $n$ & CSD \% & MUB \% & 3MUB \% & SMUB \% & 3SMUB \%\\
		\hline
		\multirow{7}{*}{2} 
		& 3 & 10.3 & $-4.5$ & $-22.8$ & 13.3 & 30.0 \\
		& 4 & 10.4 & $-3.9$ & $-12.6$ & 14.3 & 31.8 \\
		& 5 & 10.2 & $-1.5$ & $-24.2$ & 16.0 & 27.2 \\
		& 6 & 9.5 & $-4.1$ & $-23.4$ & 14.4 & 25.2 \\
		& 7 & 9.0 & $-2.5$ & $-36.6$ & 15.4 & 19.4 \\ 
		& 8 & 8.6 & $-6.7$ & $-44.4$ & 12.9 & 14.5 \\
		& 9 & 8.1 & $-6.1$ & $-61.1$ & 13.7 & 7.7 \\
		\hline
		\multirow{5}{*}{3} 
		& 3 & 37.4 & $-2.1$ & $2.7$ & 31.1 & 38.9 \\
		& 4 & 41.7 & $-8.3$ & $-25.9$ & 32.6 & 25.6 \\
		& 5 & 42.1 & 6.5 & $-19.9$ & 34.5 & 26.9 \\
		& 6 & 40.8 & 6.5 & $-30.0$ & 33.5 & 21.5 \\
		& 7 & 39.6 & 5.7 & $-38.5$ & 31.8 & - \\
		\hline
		\multirow{4}{*}{4} 
		& 3 & 35.9 & $-2.6$ & $-3.1$ & 32.0 & 33.5 \\
		& 4 & 43.4 & 5.3 & $-9.0$ & 34.7 & 28.2 \\
		& 5 & 44.4 & 9.1 & $-17.3$ & 36.3 & 23.1 \\
		& 6 & 42.4 & 9.2 & $-23.5$ & 35.0 & - \\
		\hline
		\multirow{3}{*}{5} 
		& 3 & 37.9 & $-7.4$ & $-8.9$ & 34.5 & 31.7 \\
		& 4 & 45.2 & 2.4 & $-10.0$ & 38.1 & 30.4 \\
		& 5 & 45.7 & 10.8 & $-9.1$ & 37.3 & 28.7 \\
		\hline \hline
	\end{tabular}
\end{table}

Next, we take the SD protocol as a benchmark and compare the sample efficiencies of the eight protocols mentioned above.
Suppose the SD protocol requires $N_{\SD}$ samples on average to reach a given precision, and another protocol  requires $N$ samples on average to reach the same precision. Then the reduction rate of this protocol compared with the SD protocol is defined as $(N_{\SD} - N)/N_{\SD}$. 
\Tref{tab:compareHaardn} summarizes the reduction rates of the CSD, MUB, 3MUB, SMUB,  3SMUB, SCMUB, and 3SCMUB protocols for $n$-qudit Haar-random pure states with $n,d=2,3,4,5$. Together with \fref{fig:hboxCompare}, it shows that the CSD protocol and 3SMUB protocol are the best choices with respect to the spectral gaps and sample costs. On the other hand,  the SCMUB and 3SCMUB protocols are quite appealing because they can achieve comparable performance using only $\caO(n)$ distinct tests. 

Next, we provide a bit more information on the CMUB and 3CMUB protocols, which seem to be the most elusive. \Fref{fig:GapInvCMUB3CMUBlog2} shows the inverses of average spectral gaps  achieved by the two protocols applied to Haar-random pure states. When $d=2$, the average spectral gap achieved by the CMUB protocol decreases exponentially with $n$. In all other cases, by contrast, the average spectral gap tends to converge to a positive constant determined by the local dimension $d$. For the 3CMUB protocol, the average spectral gap may have a nontrivial universal lower bound even when $d=2$. The reasons behind these intriguing phenomena merit further exploration.

\subsection{Comparison for Dicke states} 

Here we compare the verification efficiencies of the SD, CSD, MUB, 3MUB, SMUB, and 3SMUB protocols applied to Dicke states. Note that the CMUB, 3CMUB, SCMUB, and 3SCMUB protocols cannot verify all Dicke states (see \tref{tab:compareHuang} in \aref{app:compareHPS} below). Sample reduction rates mirror the counterparts for Haar-random pure states, but with  sample costs averaged over all Dicke states $|\rmD_d^n(\bfn)\>$ with  a given local dimension $d$ and qudit number $n$, where $\bfn$ satisfies the conditions in \eref{eq:DickenCondition}. The reduction rates of the CSD, MUB, 3MUB, SMUB, and 3SMUB protocols compared with the SD protocol for  $d=2,3,4,5$ and certain values of $n$ are shown in \tref{tab:compareDickedn}. In contrast with the results on Haar-random pure states (see \tref{tab:compareHaardn}), here the 3SMUB protocol is less efficient than the SMUB protocol for some values of $n$ when $d \ge 3$. For Dicke states,  some tests featured in the 3SMUB protocol tend to reduce the efficiency and are thus not necessary.

\subsection{Comparison for W states}

\begin{figure}[!b]
	\centering
	\includegraphics[scale=0.45]{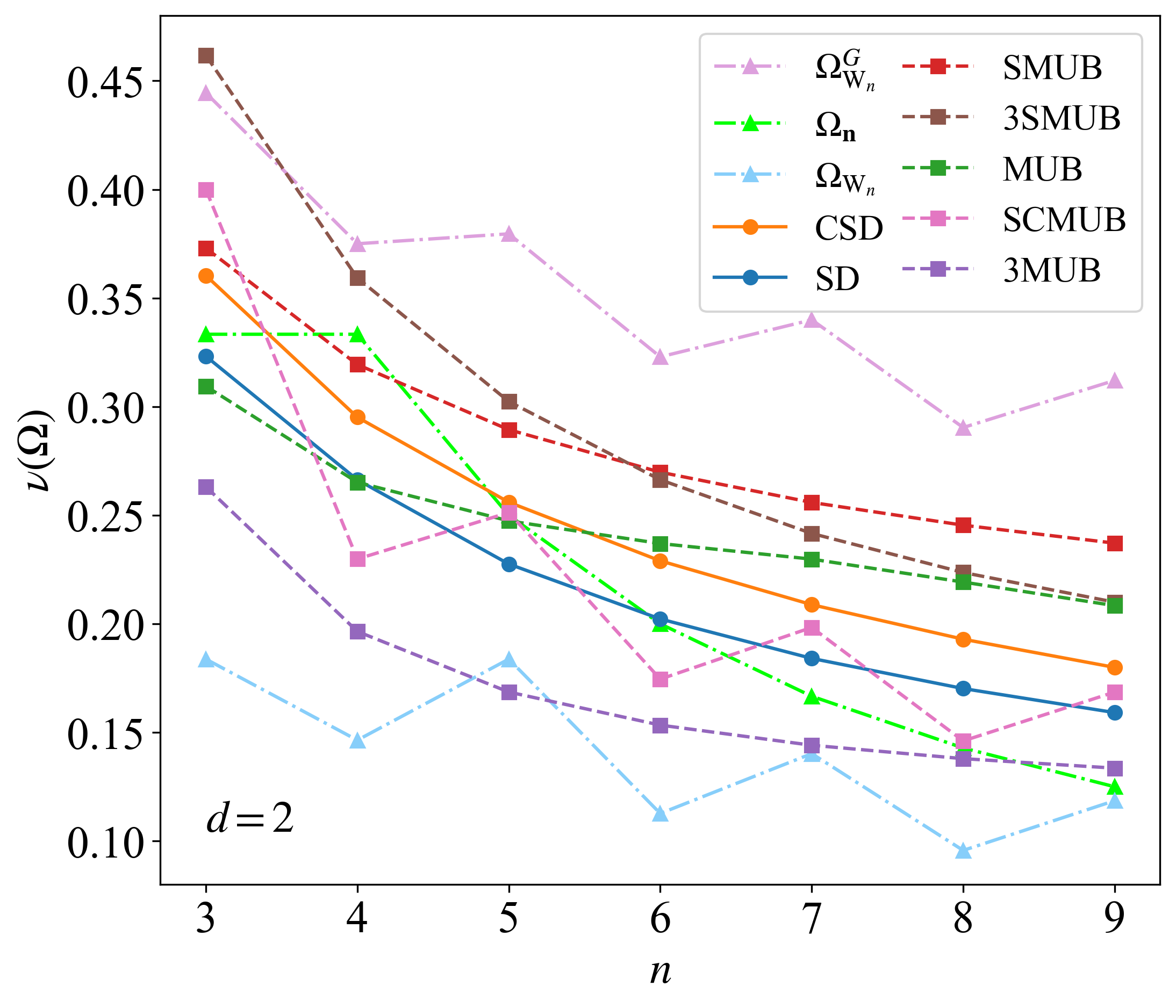}
      \vspace{-0.5ex}
	\caption{The spectral gaps achieved by the SD, CSD, MUB, 3MUB, SMUB, 3SMUB, and SCMUB protocols (under uniform probability distributions) for the $n$-qubit W state. 
As benchmarks, triangles on dash-dot lines denote the spectral gaps achieved by three specialized verification protocols 
as reproduced in 
Eqs.~\eqref{eq:Omegabfn}, \eqref{eq:OmegaWn}, and \eqref{eq:OmegaWnG}, respectively \cite{LiuYSZ19,Li2021Dicke}. }
\label{fig:WgapCompareAll}
\end{figure}

Here we focus on W states, which are a special example of Dicke states with $d=2$, $\ell=1$, and $n_1=1$ (see \aref{app:Dicke}). To be specific, the $n$-qubit W state can be expressed as follows:
\begin{equation}
    |\rmW_n\> = \frac{1}{\sqrt{n}} \sum_{\bfj \in \scrB_n^1} |\bfj\>,  \label{eq:Wstate}
\end{equation}
where $\scrB_n^1$ denotes the set of strings in $\bbZ_2^n$ with Hamming weight 1.
Previously, several efficient verification protocols for  W states were proposed in \rscite{LiuYSZ19,Li2021Dicke}, and one of the protocols can achieve the spectral gap $\nu(\Omega_\bfn)$  presented in \eref{eq:Omegabfn}. In addition, \rcite{Li2021Dicke} proposed two other protocols, which can achieve the following spectral gaps, respectively:
\begin{align}
 \nu\left(\Omega_{\rmW_n}\right) &= 
    \begin{cases}
    \frac{1}{2} - \frac{1}{\sqrt{10}} & n=3, \\[1ex]
    \frac{1 - \sqrt{1-h(n-3)}}{2} & n \ge 4,
    \end{cases} \label{eq:OmegaWn} \\[1ex]
    \nu\left(\Omega_{\rmW_n}^G\right) &= \frac{1 + (n-2)h(n-1)}{n + (n-2)h(n-1)},  \label{eq:OmegaWnG}
\end{align}
where
\begin{equation}
     h(n) := \frac{1}{2^n} \sum_{k=0}^{n} \frac{\binom{n}{k}}{1+(n-2k)^2}.
\end{equation}

Now, we compare the verification efficiencies of the  SD, CSD, MUB, 3MUB, SMUB, 3SMUB, and SCMUB protocols applied to W states. The spectral gaps achieved by these protocols (under uniform probability distributions) are shown in \fref{fig:WgapCompareAll}. When $n\geq 6$, the SMUB protocol can achieve the largest spectral gap and is thus the most efficient. As benchmarks, the figure also shows the spectral gaps achieved by three specialized verification protocols \cite{LiuYSZ19,Li2021Dicke}
as reproduced in 
Eqs.~\eqref{eq:Omegabfn}, \eqref{eq:OmegaWn}, and \eqref{eq:OmegaWnG}.

\section{Verification protocols beyond MUB} \label{app:beyondMUB}
According to the previous discussions, symmetrization and more local bases can usually increase the spectral gap, especially when the local dimension $d$ is small. In this section  we are particularly interested in verification protocols whose test numbers do not increase with the qudit number $n$, so more local bases are crucial for enhancing the efficiency.
However, each set of MUB in dimension $d$ can contain at most $d+1$ bases, so we need to consider verification protocols beyond MUB if more local bases are required. Here, we present some preliminary results for the special case $d=2$, in which the performance of the CMUB and 3CMUB protocols are not satisfactory as reflected in \fref{fig:GapInvCMUB3CMUBlog2}.

\begin{figure}[b]
	\centering
	\includegraphics[scale=0.7]{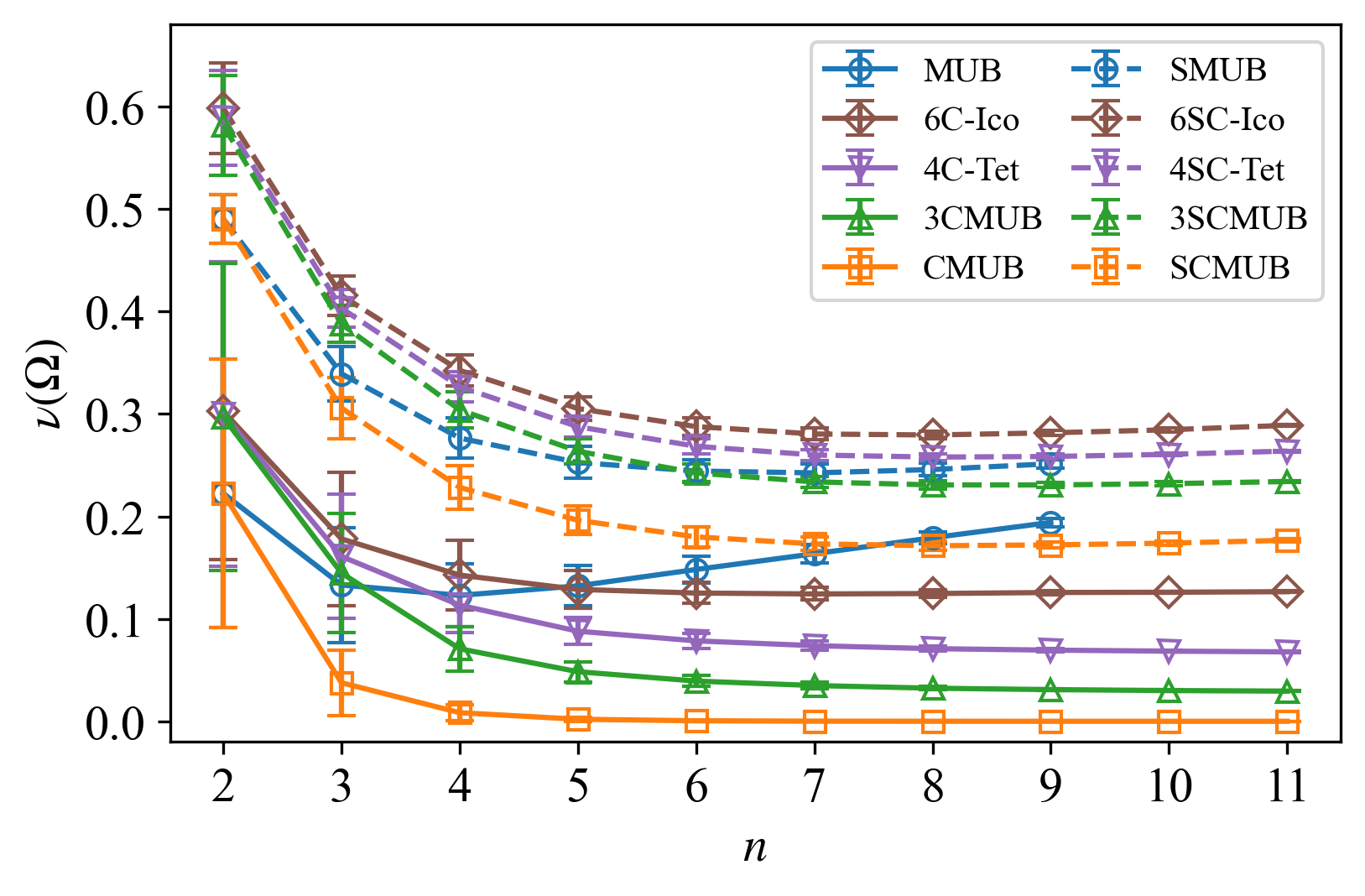}
      \vspace{-0.5ex}
	\caption{Averaged spectral gaps achieved by the 4C-Tet, 6C-Ico,  4SC-Tet, and 6SC-Ico protocols and six variants of the MUB protocol (under uniform probability distributions) for $n$-qubit Haar-random pure states. }
\label{fig:HaarPlantonic}
\end{figure}

In the case of a qubit, each Bloch vector (or point on the Bloch sphere) determines an (orthonormal) basis and the corresponding rank-1 projective measurement (note that two antipodal points determine the same basis essentially). For example, from 
the six vertices of a regular octahedron inscribed in the Bloch sphere, we can generate three bases that are mutually unbiased and thereby construct multiple variants of the 3MUB protocol. In a similar spirit, we can generate four bases from the vertices of a regular tetrahedron (Tet): 
\begin{equation}
    \bmu_1 = \frac{1}{\sqrt{3}} (1, 1, 1), \quad   \bmu_2 =  \frac{1}{\sqrt{3}} (1, -1, -1), \quad
    \bmu_3 =\frac{1}{\sqrt{3}} (-1, 1, -1), \quad \bmu_4 =\frac{1}{\sqrt{3}} (-1, -1, 1).
\end{equation}
By virtue of these bases and correlated measurements 
we can construct the 4C-Tet and 4SC-Tet protocols in analogy to the 3CMUB and 3SCMUB protocols. Likewise, we can generate six bases from six of the twelve vertices of a regular icosahedron (Ico):
\begin{equation}
\begin{aligned}
    \bmv_1 &= \frac{1}{\sqrt{1+g^2}} (1, g, 0), \quad &&\bmv_2 =  \frac{1}{\sqrt{1+g^2}} (-1, g, 0), \quad &&\bmv_3 = \frac{1}{\sqrt{1+g^2}} (0, 1, g), \\
    \bmv_4 &= \frac{1}{\sqrt{1+g^2}} (0, -1, g), \quad &&\bmv_5 =  \frac{1}{\sqrt{1+g^2}} (g, 0, 1), \quad &&\bmv_6 = \frac{1}{\sqrt{1+g^2}} (g, 0, -1),    
    \end{aligned}
\end{equation}
where $g=\left(1+\sqrt{5}\lsp \right)/2$.
By virtue of these bases and correlated measurements  we can construct the 6C-Ico and 6SC-Ico protocols. Note that the four protocols 4C-Tet, 6C-Ico, 4SC-Tet, and 6SC-Ico  consist of 4, 6, $4n$, and $6n$ distinct tests, respectively.

\Fref{fig:HaarPlantonic} illustrates the average spectral gaps  achieved by the four  protocols constructed above for Haar-random  pure states in comparison with six variants of the MUB protocol. Increasing the number of local bases tend to increase the average spectral gap as expected. 
For the CMUB, 3CMUB, and 4C-Tet protocols, the average spectral gaps decrease monotonically with $n$; for the other seven protocols, by contrast, the average spectral gaps first decrease but eventually increase with $n$, 
although the variations are very small when $n\geq 7$. 
Notably, the 6C-Ico protocol can verify Haar-random pure states with a constant number of distinct tests and a constant sample cost, independent of the qubit number $n$. 
In addition, the 6SC-Ico protocol is the most efficient among the ten protocols, although its test number increases only linearly with $n$, in  sharp contrast with the MUB and SMUB protocols.

All the protocols constructed above are tied to 2-designs. Previously, 2-designs also played an important role in constructing nearly optimal protocols for verifying bipartite pure states \cite{LiHZ19}, although they are not crucial for establishing the main results in this work. 
When the local dimension is a prime power, 2-designs can be constructed from complete sets of MUB. In general, we need to go beyond MUB; some alternative construction methods can be found in \rscite{Roy2007,GonzalezAvella2025cyclicmeasurements,Iosue2024projectivetoric}. In the future, it would be interesting to explore the potential applications of such alternative 2-designs in the verification of multipartite pure states.

\section{Comparison with the HPS protocol} \label{app:compareHPS}

Recently, HPS proposed a protocol that can certify almost all $n$-qubit pure states using single-qubit measurements~\cite{huang2025certifying}. In this Protocol (Level-$m$), all $n$ parties except for $m$ parties (chosen uniformly at random) perform Pauli $Z$ measurements, then each of the remaining parties  performs one of the three Pauli measurements ($X$, $Y$, and~$Z$) randomly. After $N$ runs, one can compute the shadow overlap based on the measurement outcomes. The state prepared is accepted if the shadow overlap is larger than a threshold and rejected otherwise. To verify an $n$-qubit pure state $|\Psi\>$ within infidelity $\varepsilon$ and significance level $\delta$, the sample cost (number of measurements) required reads  $T=2^{2m+4} \frac{\tau^2}{\varepsilon^2} \ln(\frac{2}{\delta})$, where $\tau$ is the relaxation time of a Markov chain determined by the target state $|\Psi\>$. 
If $|\Psi\>$ is a Haar-random pure state, then $\tau= \caO(n^2)$ with probability exponentially close to 1.

\begin{figure}[!b]
	\centering
	\includegraphics[scale=0.7]{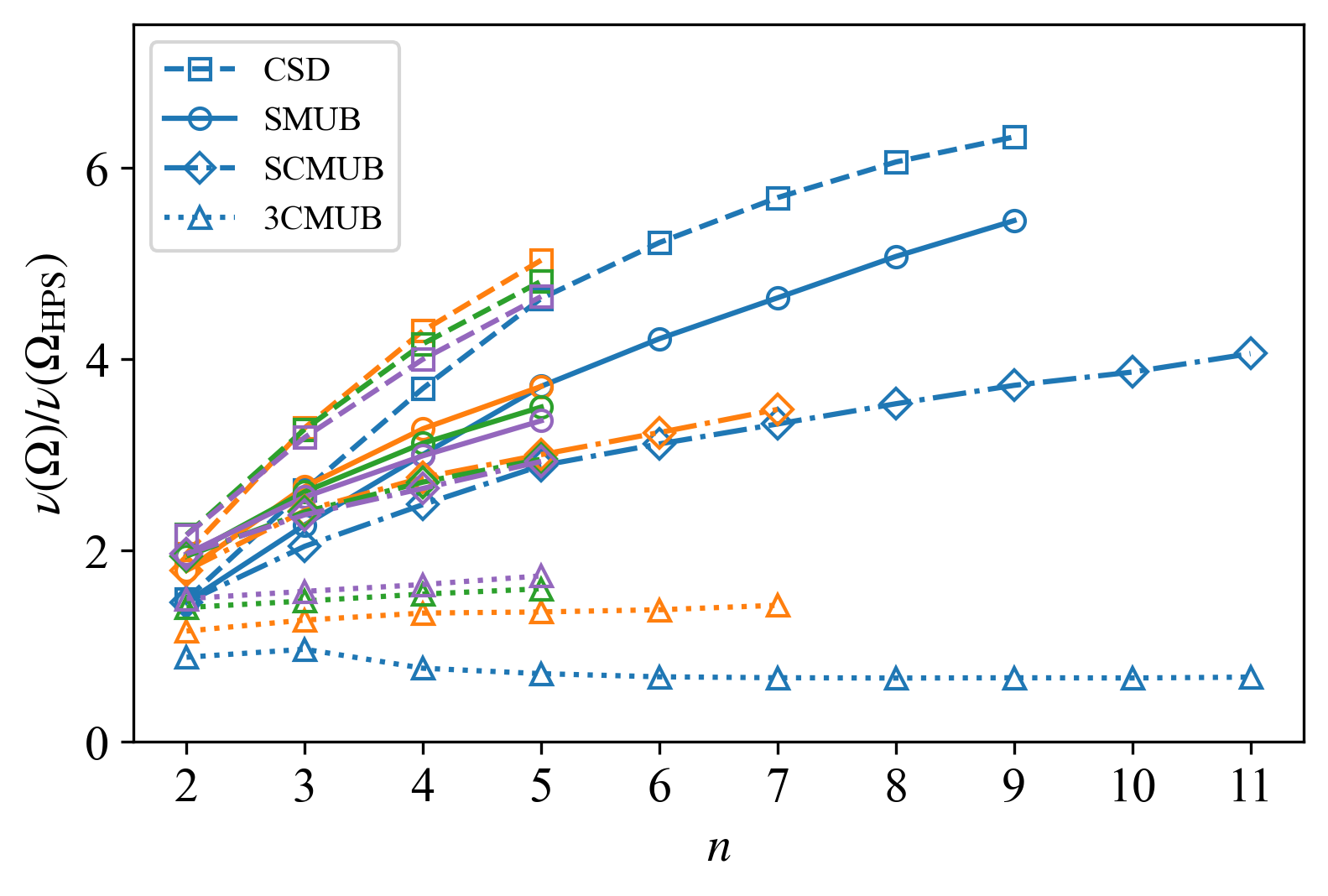}
      \vspace{-0.5ex}
	\caption{Ratios of the average spectral gaps achieved by the CSD, SMUB, SCMUB, and 3CMUB protocols  over the counterpart by the HPS protocol for $n$-qudit Haar-random pure states with $d=2,3,4,5$. The color encoding of the local dimension $d$ follows \fref{fig:HaarMUB} in the main text (blue, orange, green, and purple mean $d=2,3,4,5$, respectively).} 	\label{fig:GapRatioHPS}
\end{figure}

In general, the HPS protocol  cannot be formulated in the framework of standard QSV considered in this work. Nevertheless, HPS also proposed a variant of the Level-1 protocol ($m=1$), which is compatible with the framework of standard QSV \cite{li2025new}:  all $n$ parties except for one party (chosen uniformly at random) perform Pauli $Z$ measurements, then the last party performs the projective measurement onto the conditional reduced state of the target state. In this way, the sample cost can be reduced to $T=32 \frac{\tau}{\varepsilon} \ln(\frac{1}{\delta})$; here $1/\tau$ corresponds to the spectral gap in the framework of standard QSV.

\begin{table}[tb]
	\renewcommand{\arraystretch}{1.2}\centering
	\caption{Comparison of the SD, MUB, SMUB, 3CMUB, SCMUB, 6C-Ico, and HPS protocols with respect to the scope of applications, number of distinct tests, and sample complexity. The CSD protocol is comparable to the SD protocol except for the test number; similarly, 3MUB (3SMUB) protocol is comparable to the MUB (SMUB) protocol; 3SCMUB protocol is comparable to the SCMUB protocol; CMUB is comparable to the 3CMUB protocol except when $d=2$.}   \label{tab:compareHuang} 
	\begin{tabular}{c|cccc  ccc}
	\hline \hline 
	& SD & MUB & SMUB & 3CMUB & SCMUB & 6C-Ico & HPS \\
	\hline 
	Scope \rule{0pt}{5ex}  & \parbox{1.8cm}{\centering all $n$-qudit pure states} &  \parbox{1.5cm}{\centering$n$-qudit pure states} &  \parbox{1.5cm}{\centering$n$-qudit pure states} &  \parbox{1.5cm}{\centering$n$-qudit pure states} &  \parbox{1.5cm}{\centering$n$-qudit pure states} &  \parbox{1.5cm}{\centering$n$-qubit pure states} &  \parbox{1.9cm}{\centering almost all $n$-qubit pure states}\\[3ex]
	Test number & $2^{n-1}$ & $2^{n-1}$ & $ 2^{n-1}n$ & $3$ & $2n$ & $6$ & $n$ \\
	GHZ states   & yes & yes & yes & yes & yes & yes & no \\
	Dicke states  & yes & yes & yes & most & most & yes & no  \\[0.5ex]
	Worst case   & $\caO(2^n/\varepsilon)$ & - & - & $\infty$ & $\infty$ & -  &  $\infty$\\[1ex]
	\parbox{2.1cm}{\centering Haar-random (rigorous)}    &  $\caO(2^n/\varepsilon)$ & - & - & - & - & - &   $\caO(n^2/\varepsilon)$ \\[1.6ex]
	\parbox{2.1cm}{\centering Haar-random (numerics)}  & $\caO(1/\varepsilon)$ & $\caO(1/\varepsilon)$ & $\caO(1/\varepsilon)$ & $\caO(n/\varepsilon)$ & $\caO(1/\varepsilon)$ & $\caO(1/\varepsilon)$  &  $\caO(n/\varepsilon)$ \\[1ex]	
	\hline \hline 
\end{tabular}

\vspace{5ex}
\caption{Numbers of  $n$-qudit Haar-random pure states sampled in numerical simulations for various verification protocols proposed in this work and the HPS protocol.} \label{tab:SampleNumHaar}
	\begin{tabular}{c|c| ccc ccc}  
		\hline \hline 
	$d$ \rule{0pt}{4ex} & $n$ & SD & CSD & \parbox{2.5cm}{\centering MUB, 3MUB, SMUB, 3SMUB} & \parbox{3cm}{\centering CMUB, 3CMUB, SCMUB, 3SCMUB} & \parbox{2.8cm}{\centering 4C-Tet, 4SC-Tet, 6C-Ico, 6SC-Ico} & HPS \\[1.5ex] 
		\hline
		\multirow{4}{*}{2} & 2-8 & 1000 & 1000 & 1000 & 1000 & 1000 & 1000 \\
		& 9 & 100 & 100 & 100  & 100 & 100 & 100 \\
		& 10 & 100 & 100 & -  & 100 & 100 & 100 \\
		& 11 & - & - & - & 100  & 100 & 100 \\
		\hline
		3, 4, 5 & 2-5  & 1000 & 1000 & 1000 & 1000  & - & 1000\\
		\hline
		3 & 6, 7  & - & - & - &  100 &  - & 100 \\
		\hline
		6, 7 & 2-5 & 1000 & - & - & -& - & - \\
		\hline
		\multirow{2}{*}{8} & 2, 3, 4 & 1000 & - & - & - & - & -\\
		& 5 & 100 & -  & - & - & - & - \\
		\hline
		\multirow{2}{*}{9} & 2, 3, 4 & 1000 & - & -  & -& - & -\\
		& 5 & 20 & - & - & - & - & - \\
		\hline \hline
	\end{tabular}
\end{table}

Although the HPS protocol can certify almost all $n$-qubit pure states with $\caO(n^2)$ samples, it cannot certify certain important multiqubit pure states directly, such as  GHZ states and Dicke states, because of the divergence of the relaxation time $\tau$. 
In the perspective of standard QSV, this means the spectral gap vanishes. Consider the GHZ state $|\rmG_d^n\>$ defined in \eref{eq:GHZ} for example. The $n$ test projectors and the verification operator featured in the HPS protocol are all equal to 
\begin{equation}
	\Omega= \sum_{i\in \bbZ_d} (|i\>\<i|)^{\otimes n},
\end{equation}
so the spectral gap is equal to zero. In this special case, nevertheless, this problem can be resolved by modifying the measurement bases, using the Hadamard gate for example.
Next, consider the Dicke state $|\rmD_d^n(\bfn)\>$ defined in \eref{eq:DickeState}. Now, the $n$ test projectors and the verification operator featured in the HPS protocol are all equal to 
\begin{equation}
	\Omega =\sum_{\bfj \in  \scrB(\bfn)} |\bfj\>\<\bfj|,
\end{equation}
so the spectral gap is also equal to zero. It is possible to increase the spectral gap by modifying the measurement bases; however, a universal nontrivial lower bound for the spectral gap is still not available. By contrast, the spectral gap achieved by our SD (CSD) protocol with uniform test probabilities
is lower bounded by $2^{1-n}$ for all $n$-qudit pure states (including $n$-qubit pure states).

Next, we compare the performance of the CSD, SMUB,  SCMUB, 3CMUB, and HPS protocols for $n$-qudit Haar-random pure states. The average spectral gaps achieved by these protocols are shown in \fref{fig:GapSDMUBHPS} in the main text.  Compared with the HPS protocol, the CSD, SMUB, and SCMUB protocols can achieve much larger average spectral gaps,  which are lower bounded by a universal constant according to extensive numerical calculations. 
By contrast, the 3CMUB protocol can achieve slightly smaller (larger) average spectral gaps when $d=2$ ($d\geq 3$). \Fref{fig:GapRatioHPS} further shows the ratios of the average spectral gaps achieved by the CSD,  SMUB, 3CMUB, and SCMUB protocols over the counterpart by the HPS protocol. As $n$ increases, all the ratios tend to increase (except for the one tied to the 3CMUB protocol when $d=2$), which means the sample advantages of the CSD, SMUB, 3CMUB, and SCMUB protocols become more significant.

\Tref{tab:compareHuang} summarizes the main properties of multiple variants of the  SD and MUB protocols in comparison with the HPS protocol.

\section{Sample sizes of Haar-random pure states in numerical simulations} \label{app:SampleNumHaar} 
For the convenience of the readers, in \tref{tab:SampleNumHaar} we summarize the sample sizes of Haar-random pure states employed in our numerical simulations on QSV based on various verification protocols.

\end{document}